\documentclass[a4paper,fleqn]{cas-sc}

\usepackage[numbers,sort&compress]{natbib} 

\usepackage{amssymb}
\usepackage{amsmath}
\usepackage{float}
\usepackage{url}
\usepackage{svg}
\usepackage{xfrac}
\usepackage{multicol}
\usepackage{multirow}
\usepackage{xcolor}
\usepackage{etoolbox}
\usepackage{caption}

\makeatletter
\renewcommand{\printorcid}{}
\makeatother

\def\tsc#1{\csdef{#1}{\textsc{\lowercase{#1}}\xspace}}
\tsc{TH}
\tsc{RD}
\begin{document}
\let\WriteBookmarks\relax
\def\floatpagepagefraction{1}
\def\textpagefraction{.001}
\shorttitle{Model Accuracy vs. Efficiency in Pumped Thermal Electricity Storage}
\shortauthors{T. Heo and R. Macdonald}

\title [mode = title]{Effects of charging and discharging capabilities on trade-offs between model accuracy and computational efficiency in pumped thermal electricity storage}  

% Authors
\author[1]{Taemin Heo} %% Author name
\ead{taemin@mit.edu}
\cormark[1]
\cortext[1]{Corresponding author}

\author[1]{Ruaridh Macdonald} %% Author name

\affiliation[1]{organization={MIT Energy Initiative},%Department and Organization
            addressline={77 Massachusetts Avenue, Building E19}, 
            city={Cambridge},
            postcode={02139}, 
            state={MA},
            country={USA}}

\begin{abstract}
The increasing need for energy storage solutions to balance variable renewable energy sources has highlighted the potential of Pumped Thermal Electricity Storage (PTES). In this paper, we investigate the trade-offs between model accuracy and computational efficiency in PTES systems. We evaluate a range of PTES models, from physically detailed to simplified variants, focusing on their non-linear charging and discharging capabilities. Our results show that while detailed models provide the most accurate representation of PTES operation by considering mass flow rate ($\dot{m}$) and state of charge (SoC) dependencies, they come at the cost of increased computational complexity. In contrast, simplified models tend to produce overly optimistic predictions by disregarding capability constraints. Other approximated model variants offer a practical compromise, balancing computational efficiency with acceptable accuracy. In particular, models that disregard $\dot{m}$-dependency and approximate nonlinear SoC-dependency with a piecewise linear function achieve similar accuracy to more detailed models but with significantly faster computation times. Our findings offer guidance to modelers in selecting the appropriate PTES representation for their investment models.
\end{abstract}

% Keywords
\begin{keywords}
Pumped thermal electricity storage \sep Charging and discharging capabilities \sep Energy system modeling \sep Capacity expansion modeling
\end{keywords}

\maketitle

% Main text
\section{Introduction}\label{sec:introduction}

The rapid growth of variable renewable energy (VRE) sources, such as solar and wind, presents both opportunities and challenges for the decarbonization of energy systems \cite{husin2021critical,guerra2022facing,arent2022challenges}. While these sources are essential for reducing greenhouse gas emissions, their variability introduces reliability and cost concerns for grid operators \cite{ueckerdt2015analyzing,polleux2022overview}. Energy storage will play a significant role in balancing supply and demand, enhancing grid stability, and reducing the overall costs associated with integrating large amounts of VRE into power systems \cite{mallapragada2020long,sepulveda2021design}.

Among the various energy storage options, pumped thermal electricity storage (PTES) is emerging as a particularly promising solution for long-duration energy storage (LDES). PTES (also known as a ‘Carnot battery’, ‘pumped heat electricity storage’, ‘Brayton PTES’, and ‘Joule-Brayton PTES’ in the literature) stores electricity as heat in insulated thermal reservoirs using suitable media such as solid packed beds or liquid storage tanks \cite{zhao2022thermo}. During the charging phase, an electrically driven heat pump delivers heat to a hot store, while during the discharging phase, a heat engine converts the stored heat back into electrical energy \cite{mctigue2022techno}. The thermodynamic and electromechanical principles underlying PTES technology are well-established and reliable, with numerous demonstration systems currently under development. One example commissioned by Newcastle University is a grid-scale PTES demonstrator with packed beds, a nominal power capacity of 150 $\text{kW}_\text{e}$ and an energy storage capacity of 600 $\text{kWh}_\text{e}$, designed for an 8-hour storage cycle \cite{ameen2023demonstration}. Additionally, Malta Inc. \cite{malta} is commercially developing a 100 MW grid-scale PTES system, based on the concept proposed by Laughlin \cite{laughlin2017pumped}, which uses molten salt and coolant reservoirs to support storage cycles ranging 8-200+ hours. PTES is emerging as a competitive alternative to pumped hydro energy storage due to its reduced geographical constraints while still having a long operational life and low cost per kWh \cite{smallbone2017levelised,olympios2021progress,frate2021energy,sharma2023pumped}. Moreover, PTES offers the advantage of sector coupling, enabling the transfer of surplus energy from VRE sources to residential heating, cooling, and industrial heating sectors \cite{frate2017novel,steinmann2019pumped,walter2020techno}. This capability avoids the inefficiencies of converting electricity to a stored energy and then back to electricity for heating or cooling, positioning thermal storage as a cost-effective and efficient solution for large-scale deployment.

Accurately modeling PTES systems is essential for determining their value within grid systems \cite{levin2023energy,calero2022review}. Capacity expansion models (CEMs) optimize the design of electricity grids given cost and performance details of generation, storage, and transmission technologies; emission limits and other policies; and electricity demand time series \cite{sepulveda2018role}. CEMs have been used to study the value of PTES \cite{ghilardi2024brayton} and LDES \cite{sepulveda2021design}. However, the results of CEM optimizations are contingent on the available technologies being described accurately. Current descriptions of thermal storage in CEMs are very simple, only differentiating technologies based on their cost, leakage rates, and exergetic efficiency. While these factors are important, they do not fully capture the range of operational constraints that can limit thermal storage performance. In addition, current models use the same (limited) operational constraints for all thermal storage technologies. This makes it impossible to evaluate the benefits of a thermal storage technology with higher costs but greater operational flexibility.

CEMs have used these simple models of thermal storage thus far because the important additional details are non-linear and most CEMs are linear programs of mixed-integer linear programs. Detailed non-linear numerical models of PTES with packed beds have been developed which take into account detailed heat transfer between the working fluid and bed material \cite{desrues2010thermal,white2011loss,white2013thermodynamic,white2014wave,mctigue2015parametric,guo2016performance,mctigue2016analysis,white2016analysis,mctigue2018performance}. Based on these models, techno-economic analyses of different PTES system variants have been conducted, focusing on optimizing system design, evaluating cost-effectiveness, round-trip efficiency, and operational performance to identify the most suitable configurations for large-scale energy storage  \cite{zhao2022thermo,mctigue2022techno,frate2022techno,zhang2023parametric,parisi2024techno}. However, these non-linear models have not been incorporated into CEMs as it would make the underlying optimization non-linear and much slower to run or require smaller models. Recent research has shown that CEMs must consider years or decades of data to produce robust grid designs \cite{ruggles2024planning}, so the preference has been to use simpler technology representations and longer time series.

In this paper, we show that it is possible to incorporate non-linear details of PTES operation into linear CEMs and that to not do so misrepresents their value and role in decarbonized grids. There are several additional operational constraints which could be included in a description of PTES. One minor constraint is that PTES systems require startup time to reach operational temperatures (e.g., Malta's system requires approximately 10 minutes for start-up \cite{smith2022integration}). Station loads are also necessary to manage the mass flow rate of the working fluids and the operation of storage block segments, introducing added complexity and cost \cite{ameen2023demonstration}. 

A major constraint CEMs have yet to consider fully is how the performance of a PTES changes with its state of charge (SoC), in particular its charging and discharging capability. Here, we draw a distinction between capacity and capability, where capability is the instantaneous charging or discharging power the system is capable of while the capacity is the design maximum. In PTES, charging and discharging power depends on the temperature difference between the working fluid and the storage media. During the charging phase, hot working fluid is injected into the top of the tank, initially heating the upper storage media while the bottom remains at its starting temperature. This heat transfer process creates an axial temperature profile, with a thermal front marking the transition. As the thermal front reaches the end of the tank, the temperature difference decreases, reducing the system's charging capability. During the discharging phase, the cycle reverses: cold working fluid is injected from the bottom, the thermal front moves upward, and the discharging capability decreases as it approaches the other end of the tank. The shape of the thermal front also depends on the mass flow rate of the working fluid, and thus so does corresponding charging or discharging power. This makes the charging and discharging capabilities of PTES a non-linear function of the mass flow rate and SoC. While this complex behavior can be modeled using wave propagation and solved through the finite volume method \cite{white2011loss,white2014wave}, incorporating such detailed models into dispatch and investment optimization significantly increases computational complexity, leading to longer runtimes and memory usage. Alternative ways to represent heat transfer in PTES have been implemented in techno-economic analyses \cite{macdonaldaintegrated}, but \citet{sepulveda2021design} and \citet{ghilardi2024brayton} did not include these details in their analysis due to computational limitations. To our knowledge, no CEMs in the literature combine charging and discharging capabilities of PTES with a high temporal resolution and long horizons.

In this paper, we developed several increasingly accurate representations of PTES for use in CEMs and examined the trade-offs between their runtime and accuracy. The computational modeling community has addressed these trade-offs and challenges in representations of other generation and storage technologies \cite{abdelwahab2023daylighting,fotouhi2016study,khalik2019trade,fattahi2021measuring}. For instance, \citet{falth2023trade} investigated hydropower models with varying levels of physical and technological detail. By exploring different modeling approaches for charging and discharging capabilities of PTES, we aim to provide insights into the minimum technical detail PTES models must have to give credible results and strategies to balance model accuracy with practical usability for modellers who desire additional detail. This work has the following research aims:

\begin{enumerate}
\item To examine how much a commonly simplified PTES model deviates from more physically accurate models that account for mass flow rate and SoC dependencies in charging and discharging capabilities.
\item To explore the trade-offs between accuracy and computational time in different PTES models, with a focus on various approaches for modeling charging and discharging capabilities.
\item To evaluate the scalability of the models when integrated with a larger investment model, utilizing the GenX framework \cite{jenkins2017enhanced,genx}.
\end{enumerate}
\section{Method}\label{sec:method}
This section provides an overview of the models developed and evaluated in this study. We present the charging and discharging capabilities of PTES as nonlinear functions of mass flow rate and state of charge (SoC), along with model formulations and explanations of how each model addresses these non-linearities. Lastly, we describe the two case studies we use to study the models, the first being a price-taker optimization using selected wholesale electricity market hourly day-ahead Locational Marginal Prices (LMP), and the second a capacity expansion optimization of a three-zone electricity network.

\subsection{An overview of all models}\label{sec:2.1}
\begin{figure}[ht]
\centering
\includegraphics[width=0.8\textwidth]{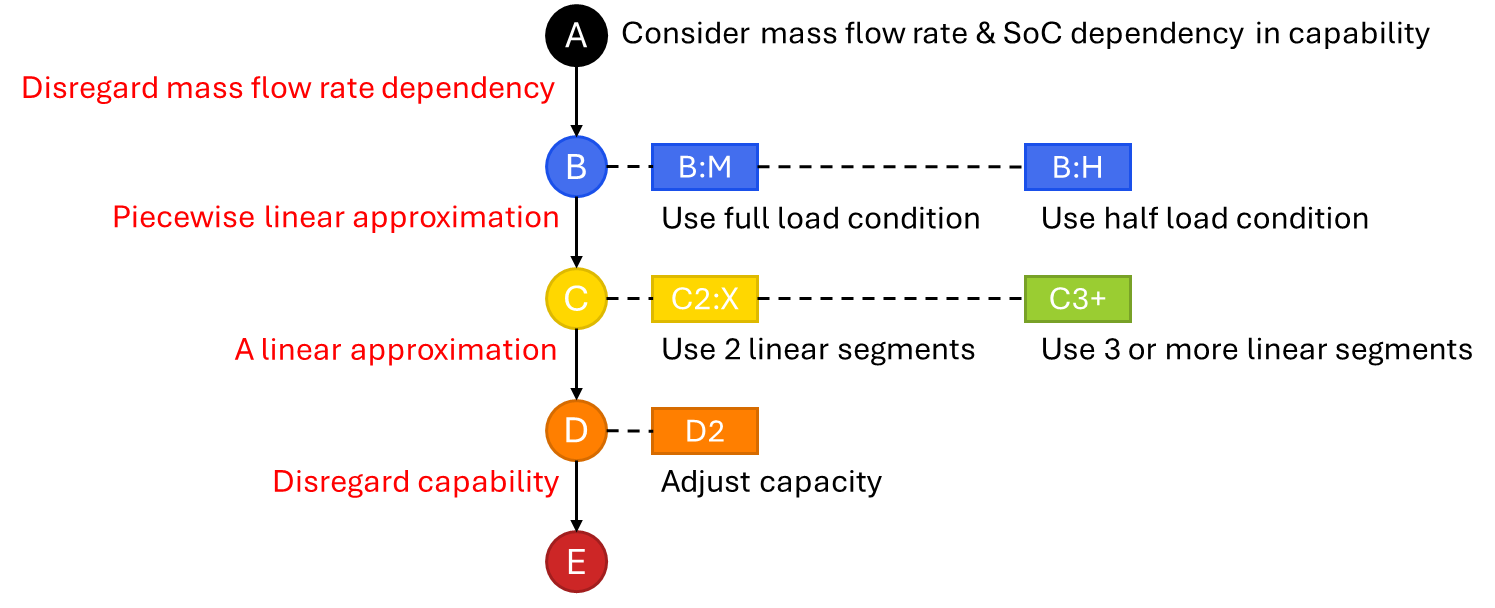}
\caption{Conceptual overview of the models from most physically accurate (Model A) to the simplest and common (Model E).}\label{fig:conceptual_overview}
\end{figure}

We developed several PTES models with varying levels of detail to represent their SoC-dependent charging and discharging capabilities. Fig.~\ref{fig:conceptual_overview} provides a conceptual overview of all the models tested in this study.

Model A is the most physically accurate, accounting for both mass flow rate ($\dot{m}$) and SoC dependencies. The Model B variants disregard the $\dot{m}$-dependency. The Model C variants approximate Model B's capability function using piecewise linear functions, enabling the optimal dispatch problem to be solved with linear solvers. The Model D variants simplify the capability function to a linear approximation. Lastly, Model E is the most simple yet commonly used model which assumes charging and discharging are not SoC-dependent and the PTES charging and discharging capabilities are always equal to the corresponding nameplate capacities. All of the models are implemented in the Julia programming language using the JuMP package \cite{Lubin2023}, which facilitates their integration with the established electricity resource capacity expansion model GenX \cite{genx}. 

The Gurobi solver \cite{gurobi2022} is employed to solve Models C, D, and E variants, while the Ipopt solver \cite{wachter2006implementation} is used for Models A and B variants, both with their default settings.

\subsection{Calculating the PTES capability functions}\label{sec:2.2}
All our PTES models, except Model E, incorporate charging and discharging capabilities functions, $\eta(\cdot)$. In those models, a PTES system with nameplate charging and discharging capacities $\overline{W}_{ch}$ and $\overline{W}_{dis}$ is only able to charge or discharge up to $\eta_{ch}(\cdot)\overline{W}_{ch}$ and $\eta_{dis}(\cdot)\overline{W}_{dis}$ respectively. These capability functions depend on one or more state and operating variables of a PTES. Calculating a capability function should be done using detailed physical models of the PTES in question and is a separate process from finding approximations of the capability functions which allow the PTES to be incorporated into a CEM.

\begin{figure}[ht]
\centering
\includegraphics[width=0.8\textwidth]{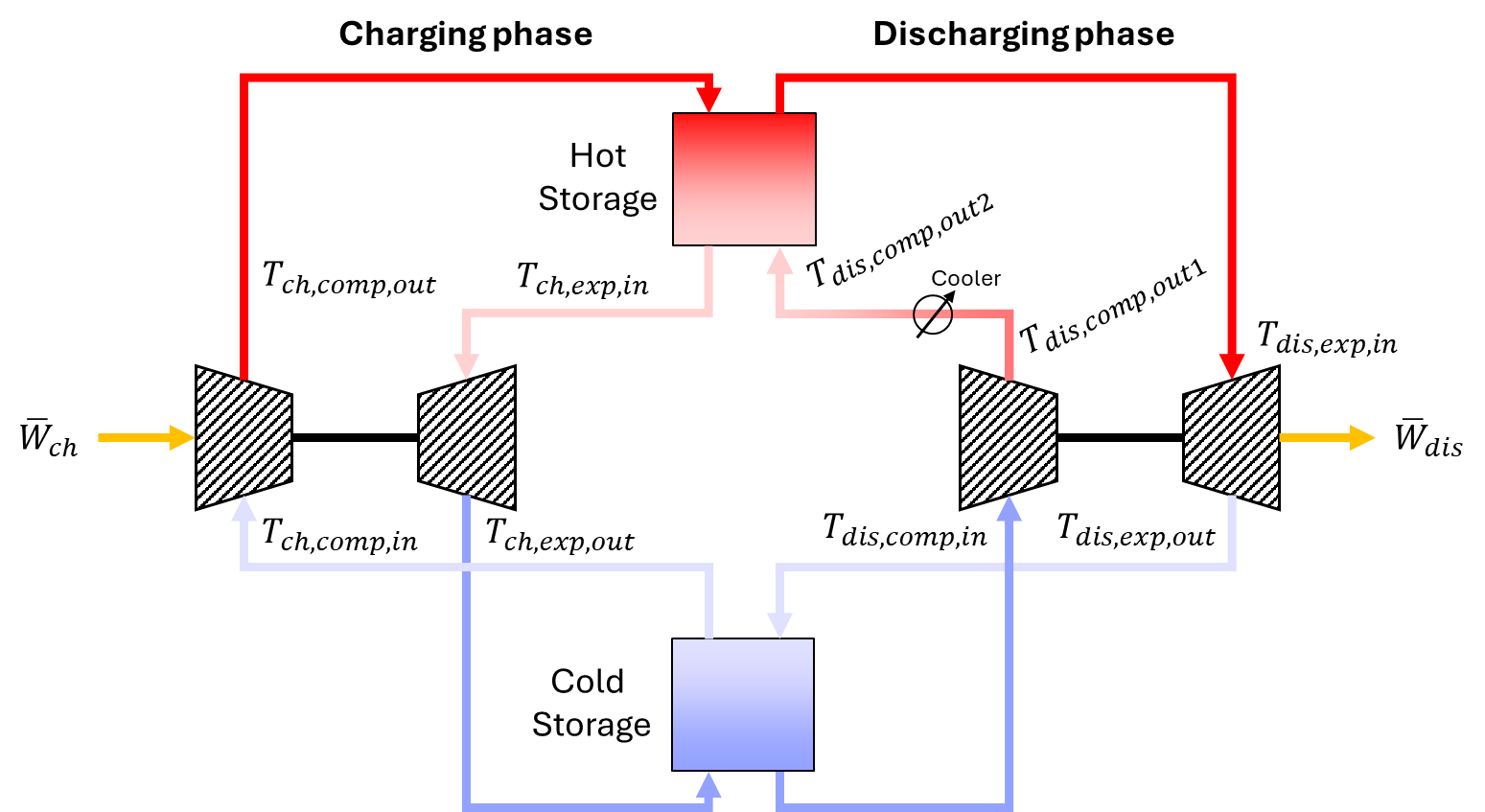}
\caption{Charging and discharging cycles of a PTES system. Operational temperatures are annotated on the flow diagram of the working fluid.}\label{fig:ptes}
\end{figure}

The thermodynamic cycles operated by the heat pump and engine of a PTES are influenced by factors such as pressure ratio, polytropic efficiencies of the compressor and expander, thermal properties of the working fluid and storage media, as well as other heat loss factors. For this paper, we use a PTES system with two packed beds, combining a mathematically optimized model developed by \citet{frate2022techno} and the commissioned system demonstrated by \citet{ameen2023demonstration}. Fig.~\ref{fig:ptes} schematically illustrates the charging and discharging cycles of the PTES system, and the detailed system configuration is provided in the Supplementary material.

We take the thermodynamic cycle and corresponding operational temperatures of each component to be fixed, so are not part of the optimization. This allows us to determine a coefficient of performance (COP), which is the ratio between the electrical work done by the compressor and expander and the thermodynamic work of heating or cooling the hot storage, as described in Eq.~\ref{eq:cop} \cite{white2009thermodynamic,ghilardi2024brayton}.
\begin{equation}
\begin{split}
\alpha_{ch} &= \frac{\overline{Q}_{ch}}{\overline{W}_{ch}} = \frac{c_p(T_{ch,comp,out} - T_{ch,exp,in})}{c_p(T_{ch,comp,out} - T_{ch,comp,in}) - c_p(T_{ch,exp,in} - T_{ch,exp,out})}\\
\alpha_{dis} &= \frac{\overline{Q}_{dis}}{\overline{W}_{dis}} = \frac{c_p(T_{dis,exp,in} - T_{dis,comp,out2})}{c_p(T_{dis,exp,in} - T_{dis,exp,out}) - c_p(T_{dis,comp,out1} - T_{dis,comp,in})}
\end{split}\label{eq:cop}
\end{equation}

\noindent where $c_p$ [kJ/kg/K] is isobaric specific heat capacity of working fluid, and $T_{i,j,k}$ [K] denotes operational temperatures, specifying whether the process is for charging (‘ch’) or discharing (‘dis’), involves the compressor (‘comp’) or expander (‘exp’), and refers to inlet (‘in’) or outlet (‘out’) temperature.

We define the system-wide COP to be the same as the COP of hot storage. If we wanted to consider sector coupling with residential heating and cooling, as in \citet{ghilardi2024brayton}, the COP should be defined for both hot and cold storage separately. This is because the system might charge or discharge solely for cold storage, rejecting all energy from the hot storage to maintain energy and sector coupling balances. In this study, we focus on electricity-to-heat-to-electricity, so cold storage is treated as an auxiliary part of the PTES operation, where its SoC always matches that of hot storage. Our approach would also apply to sector coupling with industrial heat, where the focus is on high temperatures from hot storage, while cold storage remains auxiliary. More complex modeling, including station loads from the cold storage part, could offer a more detailed view but would increase computational demands, so are not considered in our case study.

\begin{figure}[ht]
\centering
\includegraphics[width=\textwidth]{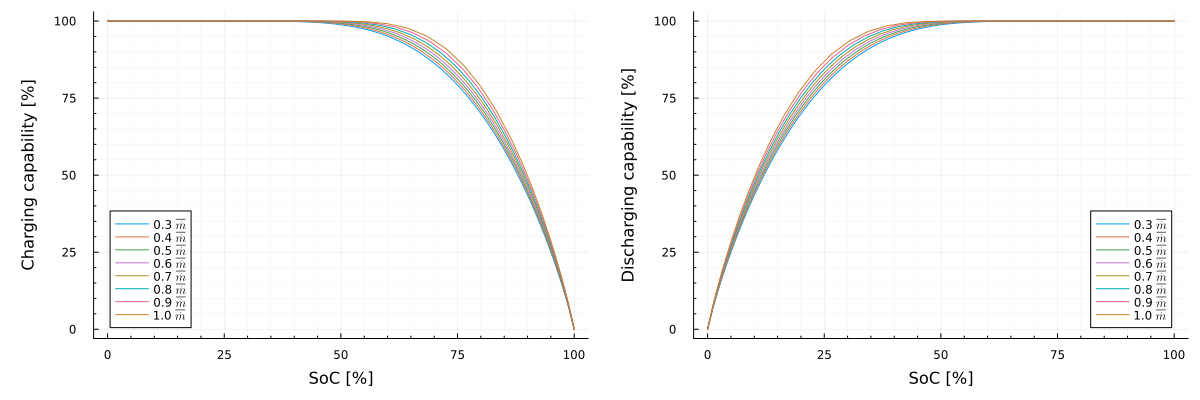}
\caption{Typical charging and discharging capability curves for a PTES system ranging from 30\% part-load to full-load conditions.}\label{fig2}
\end{figure}

Using this model, we then calculate the charging and discharging capability functions of the PTES by simulating the spatially-dependent temperature of the packed beds storages as a function of the state of charge and mass flow rate (\(\dot{m}\)) of the working fluid. The charging and discharging capability functions are shown in Fig.~\ref{fig2}. The charging capability decreases as the SoC increases because the storage media gradually becomes hotter, reducing the temperature difference between it and the working fluid and reducing the charging rate. This could be compensated for in some designs of PTES by over-sizing the working fluid pumps to further increase $\dot{m}$ but we do not consider that option here. The discharging capability behaves inversely as the storage media cools. The thermal front along the storage beds is steeper and narrower for a longer period when the PTES is operated at maximum power, and thus maximum $\dot{m}$, providing greater capability than under part-load conditions. We define the part-load level of the PTES as $p = \frac{\dot{m}}{\overline{\dot{m}}} = \frac{W_{ch}}{\overline{W}_{ch}} = \frac{W_{dis}}{\overline{W}_{dis}}.$ 

Finally, we fit a mathematical function to the numerically calculated capability functions shown in Fig.~\ref{fig2}. These function versions are used as the charging and discharging capability function of Model A, as presented in Eq.~\ref{eq:eta_A}. We describe the process of producing and fitting the capability functions in more detail in the Supplementary material.

\begin{equation}
\begin{split}
\eta_{ch}^{A}(\text{SoC},p) &= \begin{cases}1, &SoC<a_{ch}(p) \\ 1 - \left(\frac{SoC-a_{ch}(p)}{100-a_{ch}(p)}\right)^{b_{ch}(p)}, & SoC \geq a_{ch}(p)\end{cases}\\
a_{ch}(p) &= 41.4 p\\
b_{ch}(p) &= -1.683p + 5.351\\
\eta_{dis}^{A}(\text{SoC},p) &= \begin{cases}1 - \left(\frac{a_{dis}(p)-SoC}{a_{dis}(p)}\right)^{b_{dis}(p)}, & SoC < a_{dis}(p) \\ 1, &SoC\geq a_{dis}(p)\end{cases}\\
a_{dis}(p) &= -39.282 p + 100\\
b_{dis}(p) &= -1.627p + 5.373\\
\end{split}\label{eq:eta_A}
\end{equation}

\subsection{Model formulations}\label{sec:2.3}
\begin{figure}[ht]
\centering
\includegraphics[width=0.7\textwidth]{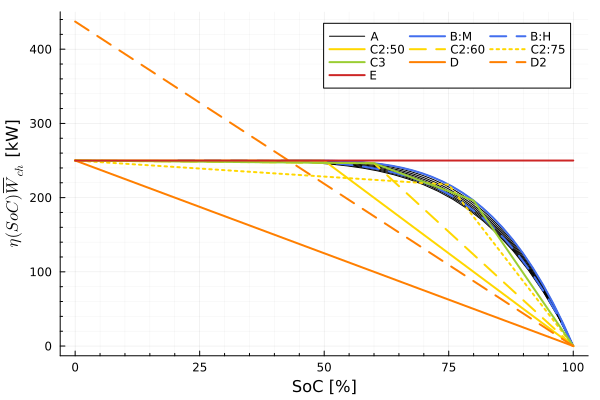}
\caption{Factored charging capabilities of a PTES system with $\overline{W}_{ch} = 250$~kW illustrating different model formulations.}\label{fig:charge_capability_models}
\end{figure}

This section provides an overview of the model formulations. For this study, the only differentiator between the PTES models is their capability functions. These are shown in Figure \ref{fig:charge_capability_models}. As such, the other constraints in our description of Model A are also true for Models B-E.

\subsubsection{Model A}\label{sec:2.3.1}

We use the COP to estimate the electricity-to-heat and heat-to-electricity conversions and to simulate the SoC of the PTES. Additionally, we assume the motor-generator has an efficiency of $\eta_{MG} = 0.98$. While the instantaneous COP under part-load conditions may deviate from the nominal value we calculated, incorporating such modeling would also add significant computational complexity so is left to future work. We introduce the variables $SoC$, $W_{ch}$, and $W_{dis}$, allowing us to express the change in SoC in hour $h\in1:H$ as follows:
\begin{equation}
    \Delta SoC_h = (\alpha_{ch} W_{ch,h} \eta_{MG} - \alpha_{dis} W_{dis,h} / \eta_{MG}) / C
\end{equation}

\noindent where $C$ [kWh] represents the energy storage capacity of the PTES. Our model accounts for energy leakage from the PTES, with an assumed hourly energy loss of $\gamma = 0.0002$, equivalent to 0.5\% energy loss per day, following references \cite{mctigue2016analysis,mctigue2022techno}. The result energy balance is:
\begin{equation}
    SoC_h = (1-\gamma) SoC_{h-1} + \Delta SoC_h
\end{equation}

To ensure no net energy gain in the PTES, we constrain the SoC at the end of the electricity price time series to revert to the initial SoC:
\begin{equation}
SoC_1 = (1-\gamma)SoC_H + \Delta SoC_1
\end{equation}

As we introduced in Sec.~\ref{sec:2.2} and Eq.~\ref{eq:eta_A}, charging and discharging power at time $h$ is bounded by the capability functions,
\begin{equation}
\begin{split}
0 \leq &W_{ch,h} \leq \eta_{ch}^A(SoC_h,p_h)\overline{W}_{ch} \\
0 \leq &W_{dis,h} \leq \eta_{dis}^A(SoC_h,p_h)\overline{W}_{dis} \\
\end{split}\label{eq:constraint_A}
\end{equation}

\noindent where $p_h = W_{ch,h}/\overline{W}_{dis}$ or $W_{dis,h}/\overline{W}_{dis}$. Since the constraints in Eq.~\ref{eq:constraint_A} is non-linear, Model A is a non-linear program. 

\subsubsection{Model B}\label{sec:2.3.2}

We define two Model B variants which disregard the $\dot{m}$-dependency. Model B:M uses the capability function corresponding to the full-load condition, while Model B:H uses the capability function for the half-load condition. Model B:M is more flexible than Model A as its capability function is equal or greater to that of Model A for all SoC, because they are not $\dot{m}$-dependent.
\begin{equation}
\begin{split}
\eta_{ch}^{B:M}(SoC) &= \eta_{ch}^{A}(SoC,1), \quad \eta_{ch}^{B:H}(SoC) = \eta_{ch}^{A}(SoC,0.5)\\
\eta_{dis}^{B:M}(SoC) &= \eta_{dis}^{A}(SoC,1), \quad \eta_{dis}^{B:H}(SoC) = \eta_{dis}^{A}(SoC,0.5)
\end{split}\label{eq:eta_B}
\end{equation}

The capability functions for the two Model B variants are given in Eq.~\ref{eq:eta_B}, and the power constraints in Eq.~\ref{eq:constraint_B} replace those in Eq.~\ref{eq:constraint_A}.
\begin{equation}
\begin{split}
0 \leq &W_{ch,h} \leq \eta_{ch}^{B:M \text{or} B:H}(SoC_h)\overline{W}_{ch} \\
0 \leq &W_{dis,h} \leq \eta_{dis}^{B:M \text{or} B:H}(SoC_h)\overline{W}_{dis} \\
\end{split}\label{eq:constraint_B}
\end{equation}

\subsubsection{Model C}\label{sec:2.3.3}

We create several Model C variants which leverage the convexity of the capability functions and objective pressure to create piecewise linear approximations of the Model B:M capability function. Each of these segments is less than or equal to the true B:M capability functions, making the Model C variants less flexible Model B:M. We divide the capability functions into $N$ segments, each defined by a linear function $\eta_n^{C\,N}(\cdot)$, indexed by $n\in1:N$. We constrain the charging and discharging power of the PTES to be less than all of these linear segments, as shown in Eq.~\ref{eq:C_ieqs}. 
\begin{equation}
\begin{split}
0 \leq W_{ch,h}& \leq \eta_{ch,n}^{C\,N}(SoC_h)\overline{W}_{ch} \quad \forall n \in 1:N \\
0 \leq W_{ch,h}& \leq \eta_{dis,n}^{C\,N}(SoC_h)\overline{W}_{dis} \quad \forall n \in 1:N
\end{split}\label{eq:C_ieqs}
\end{equation}

Note that, due to the convexity of the capability functions, $\eta_{ch,n}^{C\,N}(SoC) < \eta_{ch,n+1}^{C\,N}(SoC)$ for all $n \in 1:N-1$ if $SoC$ is less than their point of intersection; the reverse is true when $SoC$ exceeds the intersection. Consequently, for $SoC$ values below the intersection, we have $W_{ch,h} \leq \eta_{ch,n}^{C\,N}(SoC_h) \overline{W}_{ch} < \eta_{ch,n+1}^{C\,N}(SoC_h) \overline{W}_{ch}$ for all $n \in 1:N-1$, and the reverse holds when $SoC$ is greater than the intersection. This makes Eq.~\ref{eq:C_ieqs} equivalent to a single inequality constraint with a piecewise linear capability function. Our approach is preferred since formulating a piecewise linear function requires integer variables and converts the model into a mixed-integer linear program. A comparison between using multiple inequalities with linear segments (linear programs with more constraints) and a single inequality with a piecewise linear function (mixed-integer linear programs with fewer constraints) is provided in the Supplementary material.

The Model C2:X variants use two linear segments, where X denotes the breakpoint location for the charging capability function and 100-X denotes the breakpoint location for the discharging capability function, as described in Eq.~\ref{eq:eta_C2}. In our case study, we evaluate using X values of 50\%, 60\%, and 75\%.
\begin{equation}
\begin{split}
\eta_{ch,1}^{C2:X}(SoC) &= \frac{\eta_{ch}^{B:M}(X)-1}{X}SoC + 1\\
\eta_{ch,2}^{C2:X}(SoC) &= \frac{\eta_{ch}^{B:M}(X)}{100-X}(100-SoC)\\
\eta_{dis,1}^{C2:X}(SoC) &= \frac{\eta_{dis}^{B:M}(100-X)}{100-X}SoC \\ 
\eta_{dis,2}^{C2:X}(SoC) &= \frac{1 - \eta_{dis}^{B:M}(100-X)}{X}(SoC-100) + 1
\end{split}\label{eq:eta_C2}
\end{equation}

Model C3 is formulated in a similar manner, with breakpoints located at $[60\%, 80\%]$. Models with more linear segments use uniformly distributed breakpoints. We tested Models C10 and C30 in the first case study.

\subsubsection{Model D}\label{sec:2.3.4}

Model D uses a linear function to approximate the capability functions. As shown in Figure \ref{fig:charge_capability_models}, this leads to much lower capability functions than in Model A. Therefore, we create Model D2 which adjusts the nominal charging and discharging power to $\overline{W}_{ch}^{D2}$ and $\overline{W}_{dis}^{D2}$, respectively to better match the average value of Model B:M's capability functions. The cost of the nameplate capacity is also adjusted so that the capability-weighted capacities are the same.

The capability functions of Model D are $\eta_{ch}^D(SoC) = 1 - SoC/100$ and $\eta_{dis}^D(SoC) = SoC/100$. The power constraints can be expressed as:
\begin{equation}
\begin{split}
0 \leq &W_{ch,h} \leq (1 - SoC/100)\overline{W}_{ch} \\
0 \leq &W_{dis,h} \leq (SoC/100)\overline{W}_{dis} \quad \text{for Model D}\\
0 \leq &W_{ch,h} \leq (1 - SoC/100)\overline{W}_{ch}^{D2} \\
0 \leq &W_{dis,h} \leq (SoC/100)\overline{W}_{dis}^{D2} \quad \text{for Model D2}
\end{split}\label{eq:constraint_D}
\end{equation}

\subsubsection{Model E}\label{sec:2.3.5}

Lastly, Model E ignores all capability constraints, equivalent to using a constant capability function $\eta_{ch} = \eta_{dis} = 1$, as illustrated by the red line in Fig.~\ref{fig:charge_capability_models}. The power constraints are as follows:
\begin{equation}
\begin{split}
0 \leq &W_{ch,h} \leq \overline{W}_{ch} \\
0 \leq &W_{dis,h} \leq \overline{W}_{dis}
\end{split}\label{eq:constraint_E}
\end{equation}

\subsection{Price-taker case study}\label{sec:2.4}

Our first case study focuses on the optimal dispatch of PTES charging and discharging to maximize the profit $(P)$ in a price-taker system. The profit is calculated as the revenue earned from discharging (selling) power to the grid minus the cost incurred from charging (purchasing) energy to the PTES, given exogenously determined hourly electricity prices ($c_h$). The objective function is shown in Eq.~\ref{eq:objective}, where the index $h\in1:H$ corresponds to the hour of the year. It might appear that charging and discharging could occur simultaneously; however, this is not possible in practice for a PTES, as the cycle must be reversed between the two. The optimal solution of Eq.~\ref{eq:objective} naturally selects either charging or discharging, since there is no incentive to compromise revenue or increase costs by performing both simultaneously.
\begin{equation}
\begin{split}
&\max_{W_{ch,h},W_{dis,h}} P\\
P &= \sum_{h=1}^{H} (W_{dis,h} - W_{ch,h})c_h
\end{split}\label{eq:objective}
\end{equation}

For the price-taker optimizations, we fix the PTES charging, discharging and energy capacities to be 250kW, 160kW, and 11MWh respectively. More details of the design are given in the Supplementary material. Because the capacities are fixed, this made Models A and B non-linear programs and Models C-E linear. We solve the linear programs using the Gurobi solver and the non-linear programs using the Ipopt solver.

We sourced 27 one-year hourly electricity price time series for use in this case study. The U.S. Energy Information Administration provides wholesale electricity market data in the seven Regional Transmission Organizations (RTO) and Independent System Operators (ISO)~\cite{eia}. We collected hourly day-ahead LMP data from all seven RTO and ISO for the years 2020 to 2022. For notational brevity, we refer to both RTOs and ISOs as ISO, and zones and hubs as zones. We found that zone-to-zone variations in prices were relatively minor compared to year-to-year and ISO-to-ISO variations. Therefore, we selected a few representative zones from each ISO, as summarized in Table~\ref{table:ISO}. ERCOT 2021 is excluded due to extreme LMP values caused by the statewide power crisis during the winter storm. CAISO 2022 and SPP 2022 are also excluded due to missing data points in the collected LMP series. As a result, 27 price data sets are used in our case study. These hourly time series are shown in the Supplementary material.

\begin{table}[h]
\centering
\begin{tabular}{cccccc}
  \hline
  RTO/ISO & Zone or Hub & 2020 & 2021 & 2022 \\
  \hline
  CAISO   & NP15          & o    & o    & x \\
  ERCOT   & North         & o    & x    & o \\
  ISONE   & Northeast MA  & o    & o    & o \\
  MISO    & MN,IL,AR,LA   & o    & o    & o \\
  NYISO   & J-NYC         & o    & o    & o \\
  PJM     & PEPCO         & o    & o    & o \\
  SPP     & South         & o    & o    & x \\
  \hline
\end{tabular}
\caption{A summary of the collected and selected hourly day-ahead LMP data sets from representative zones of each ISO for the price-taker case study.}\label{table:ISO}
\end{table}

\subsection{Capacity expansion case study using GenX}\label{sec:2.5}

GenX is an open-source electricity resource capacity expansion model designed to support decision-making in the evolving electricity landscape. It typically uses constrained linear or mixed-integer linear optimization to determine the optimal portfolio of electricity generation, storage, transmission, and demand-side resources needed to meet future electricity demand at the lowest cost, while adhering to various operational, environmental, and policy constraints. In this work, we have extended it to non-linear optimization using versions of the Ipopt solver.

In our second case study, we use a three-zone electricity network example. This example models one year at an hourly resolution and includes zones representing Massachusetts, Connecticut, and Maine. The study involves four resources, including solar PV, onshore wind, utility-scale Li-ion battery storage, and PTES. Modifications were made to include different PTES models and to enable the use of the Ipopt solver in GenX. Notably, GenX allows charging and discharging capacities to be modeled symmetrically or asymmetrically. In symmetric mode, these capacities are identical, while in asymmetric mode, they are optimized independently. For PTES, charging and discharging capacities are fully dependent due to shared turbomachinery, where the same mass flow rate operates in opposite directions. However, the thermodynamic cycle introduces variations in the electric charging and discharging capacities from the point of view of the grid, which is how capacities are accounted in GenX. We adjusted GenX's symmetric storage formulation to account for these thermodynamic factors. Therefore, the PTES charging capacity is set at 1.5 times the discharging capacity. The Li-ion storage was modeled as symmetric storage with identical charging and discharging capacities and no capability functions.

We use GenX time domain reduction feature to select eleven one-week representative periods from our 8760-hour data set. This feature uses k-means clustering to select and weight the representative periods. The eleven periods and their weights are displayed in Fig.~\ref{fig:GenX_SoC}. Additionally, an improved formulation for long-duration storage is applied to GenX, following \citet{parolin2024improved}, to track SoC of the PTES between represenative periods.

In our capacity expansion case-study, GenX optimizes both the capacity and operation of each resource, necessitating additional input data of the capital and operating costs of PTES. We used 2022 costs under the moderate technology innovation scenario reported by \citet{parzen2023value} for PTES costs, as well as data from the National Renewable Energy Laboratory's Annual Technology Baseline \cite{atb} for all other resources. 

While all other modules of GenX are formulated as linear programs, the charging and discharging capability function introduces non-linearity due to the multiplication of $\eta(\cdot)$ and $\overline{W}_{ch}$ in the constraints. As a result, Models A to D are solved using the Ipopt solver, while Model E uses the Gurobi solver. We evaluated six models with GenX: Models A, B:M, C2:75, C3, D, and E. We found that Ipopt's default linear solver, MUMPS, often reaches its memory limit during GenX optimization. To address this, we used the MA86 linear solver and set the Ipopt parameter ‘‘mu\_strategy’’ to ‘‘adaptive,’’ following \citet{falth2023trade}’s recommendation to improve stability and solution times. However, we also observed that this setting is less effective for smaller price-taker optimizations, which is discussed in the Supplementary material. The solution times for the various models reported in this research were obtained by running the models on a 64-bit Microsoft laptop with an Intel Core i7-12700H 2.3 GHz processor and 32 GB RAM.
\section{Results}\label{sec:results}

\subsection{Price-taker case study}

First, we show how the choice of PTES model affects the operation of the PTES in our price-taker cases study. We then demonstrate that the accuracy of the optimized operations and the total runtime are inversely related.

\subsubsection{Comparison of model operations}\label{sec:3.1}

\begin{figure}[ht]
\centering
\includegraphics[width=\textwidth]{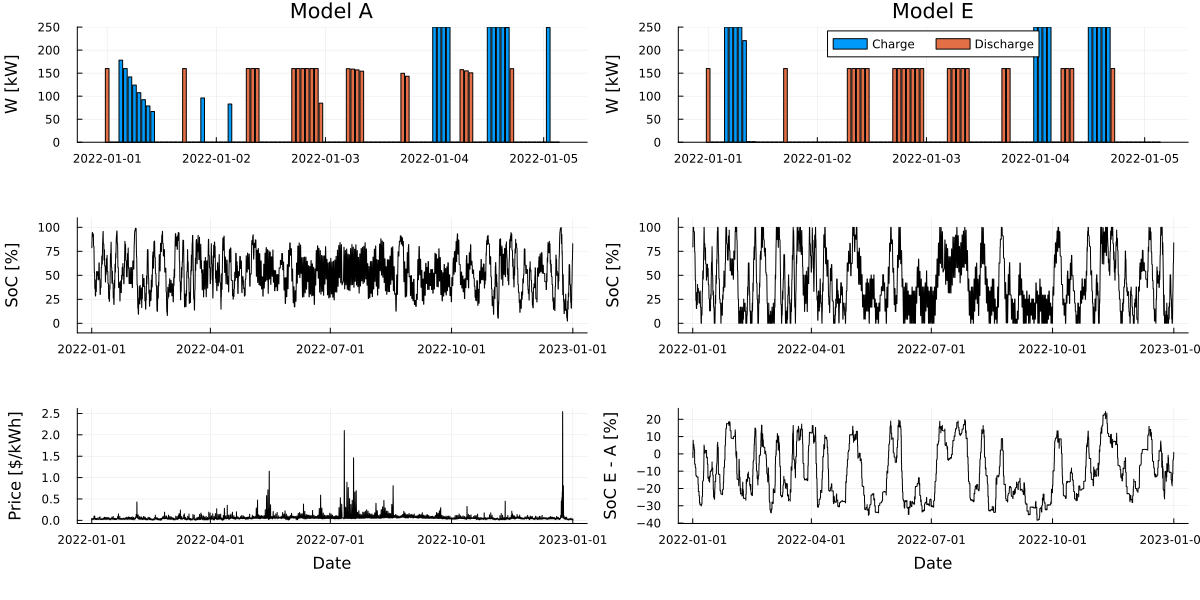}
\caption{Optimal hourly operations derived from Models A and E for the ERCOT North hub in 2022. The LMP time series is shown in the bottom left panel. The top two panels display the charging and discharging power over the first 100 hours. The two central panels present the SoC time series, and the bottom right panel shows the SoC difference between Models A and E.}\label{fig:ercot_2022_AvsE}
\end{figure}

The charging and discharging behavior of the optimized PTES operation varied significantly based on the choice of PTES model. Fig.~\ref{fig:ercot_2022_AvsE} presents the optimal hourly PTES operations under Models A and E for ERCOT North hub in 2022, with the LMP series shown in the bottom left panel. Equivalent figures for other model variants are provided in the Supplementary material.

The first few hours of charging, shown in the top row of Fig.~\ref{fig:ercot_2022_AvsE},  clearly illustrate the impact of charging capability functions. Model A's charging capability and power decrease as the PTES charges, whereas Model E allows full power charging until the PTES is fully charged. As a result of this difference, Model A spends 8 hours charging during the first day, while Model E only spends 4 hours.

\begin{figure}[ht]
\centering
\includegraphics[width=0.7\textwidth]{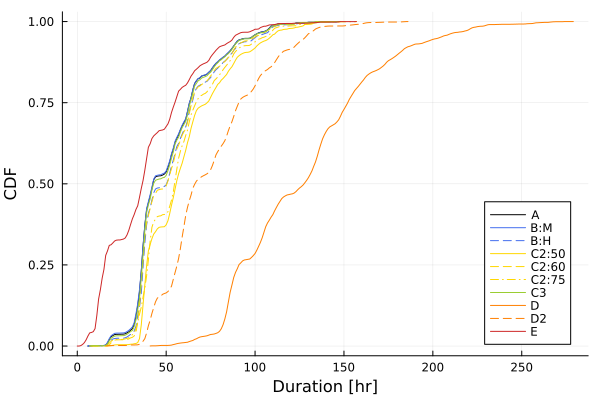}
\caption{Empirical cumulative density functions of storage durations for each model, recorded from the optimal hourly operations for the ERCOT North hub in 2022.}\label{fig:ercot_2022_duration_ecdf}
\end{figure}

The Model A PTES balances periods of charging and discharging to carefully manage its SoC, leading to less variation in SoC than under Model E. This is shown in the two central panels of Fig.~\ref{fig:ercot_2022_AvsE}. Recall that Fig.~\ref{fig:charge_capability_models} showed that the charging capability reduces much more quickly when the SoC exceeds 75\% and that the discharging capability decreases rapidly below 25\% SoC. Thus, we can see that, under Model A, the PTES operation is optimized to maintain the SoC in the region where charging and discharging capabilities are relatively high (SoC of 25 - 75\%). This is not the case for Model E, under which the PTES frequently completely empties and fills.  The full capacity is always available under Model E, enabling prompt charging and discharging regardless of SoC. The SoC under the two models differs by 20 to 30\%, implying that investment models that use Model E formulations may produce erroneous results. 
To give another perspective on how the choice of PTES model affects its operations, we analyzed how long a PTES stored energy for. We do this by tracking the number of hours that each watt-hour of energy is stored in a PTES by monitoring the time at which it is charged and discharged. We do this accounting on a first-in first-out basis, producing empirical cumulative density functions (ECDF) of the storage durations for each case. Fig.~\ref{fig:ercot_2022_duration_ecdf} shows the ECDF for all of the models in the ERCOT case. ECDFs of all 27 price data sets are shown in the Supplementary material.

Figure~\ref{fig:ercot_2022_duration_ecdf} shows that PTES stores energy for less time under the less operationally constrained models. Model E is the most flexible, able to respond to price fluctuations with maximum power. As a result, more than 25\% of its energy cycles through the storage daily. In contrast, for all other models, most energy remains stored for more than two days, as prompt charging and discharging are less effective due to capability constraints. Model B:M exhibits energy storage duration distribution that is slightly more similar to Model A than Model B:H. This can be attributed to the optimized Model A PTES operating at full-load most of the time, making Model B:M's a better approximation than Model B:H. 

The Model C2 variants highlight the importance of choosing the breakpoint location for piecewise linear approximations. The ECDF of Model C2:75 is almost identical to that of Model B:H, whereas Models C2:50 and C2:60 show longer energy storage durations. These results align with Fig.~\ref{fig:charge_capability_models}, as Model C2:75's capability function differs less from Models A and B variants. 

A clear pattern emerges: greater capability allows for more flexible operation, while more restrictive capability functions lead to longer energy storage durations. Model C2:50 has lower capability at high SoC compared to Model C2:60. We can roughly order the models by their capability at higher SoC (e.g., 80\%), as follows: E $>$ B:M $>$ C3 $>$ A $>$ B:H $>$ C2:75 $>$ C2:60 $>$ C2:50 $>$ D2 $>$ D. The ECDF also shifts to the right in the same order in Fig.~\ref{fig:ercot_2022_duration_ecdf}. 

Since the piecewise linear approximation of Model C3 is nearly identical to Model B:M, its energy storage duration distribution is also almost identical to Model B:M. Model D has the longest energy storage duration because it is the most restrictive model. As with other cases, Model D2 is less restrictive so its storage duration ECDF is less of an outlier.

\subsubsection{The trade-off between accuracy and computational time}\label{sec:3.2}

While using a more detailed PTES model improves the fidelity of the results, it also increases the runtime of the model as they require additional linear and non-linear constraints. In this section, we demonstrate how incorporating detailed charging and discharging capability functions affects computational time while improving accuracy. We use the root-mean-square deviation (RMSD) as our accuracy metric, comparing the results of each model with those using Model A. Since energy system operators and researchers may prioritize accuracy in charging and discharging power and/or SoC, we compare the RMSD for both measures across the models, as defined in Eq.~\ref{eq:rmsd}.
\begin{equation}
\begin{split}
RMSD_{SoC}^{X} &= \sqrt{\frac{\sum_{h=1}^H (SoC_h^X - SoC_h^A)^2}{H}}\\
RMSD_{W}^{X} &= \sqrt{\frac{\sum_{h=1}^H (\frac{100}{\overline{W}_{ch}}(W_{ch,h}^X - W_{ch,h}^A) + \frac{100}{\overline{W}_{dis}}(W_{dis,h}^X - W_{dis,h}^A))^2}{H}}
\end{split}\label{eq:rmsd}
\end{equation}

\noindent where $X$ denotes the Model name. Charging and discharging powers are normalized by nominal capacity and multiplied by 100 to match their RMSD units with SoC RMSD [\%].

\begin{figure}[ht]
\centering
\includegraphics[width=\textwidth]{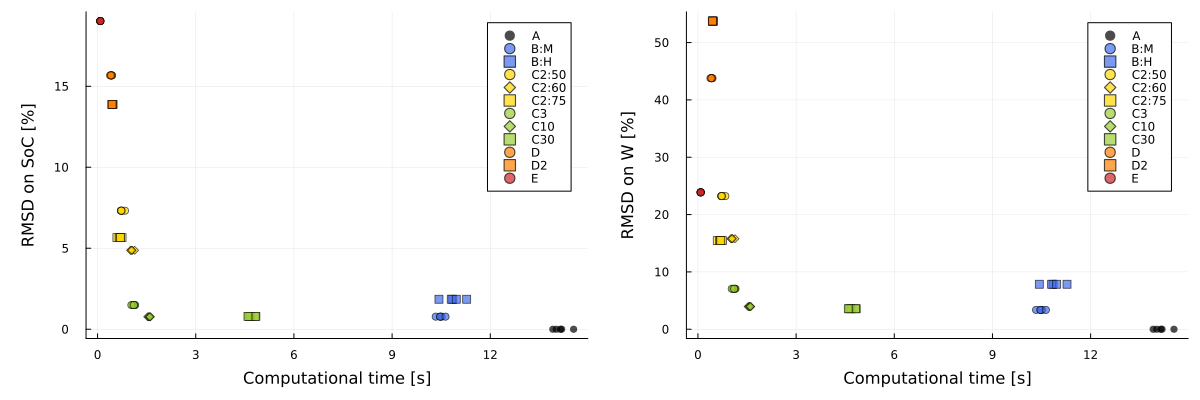}
\caption{Trade-offs between computational time and RMSD on SoC and W measured from the optimization results for the ERCOT North hub in 2022.}\label{fig:ercot_2022_duration_tradeoff}
\end{figure}

Figure~\ref{fig:ercot_2022_duration_tradeoff} illustrates the trade-offs between computational time and two RMSDs, measured from the optimization results for ERCOT North hub in 2022. Each model was run five times to account for aleatoric uncertainty in computational time. Fig.~\ref{fig:ercot_2022_duration_tradeoff} shows a clear Accuracy = 1/Speed trade-off.
 
\begin{figure}[ht]
\centering
\includegraphics[width=\textwidth]{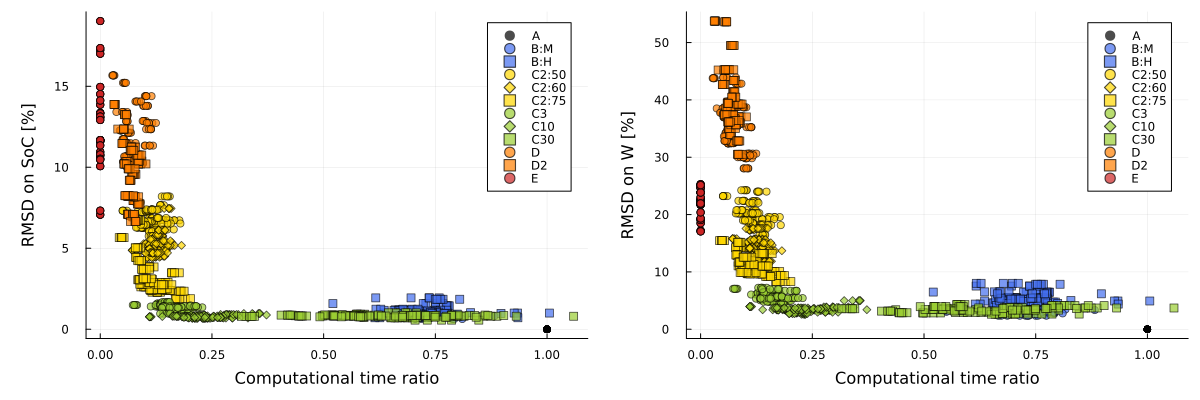}
\includegraphics[width=\textwidth]{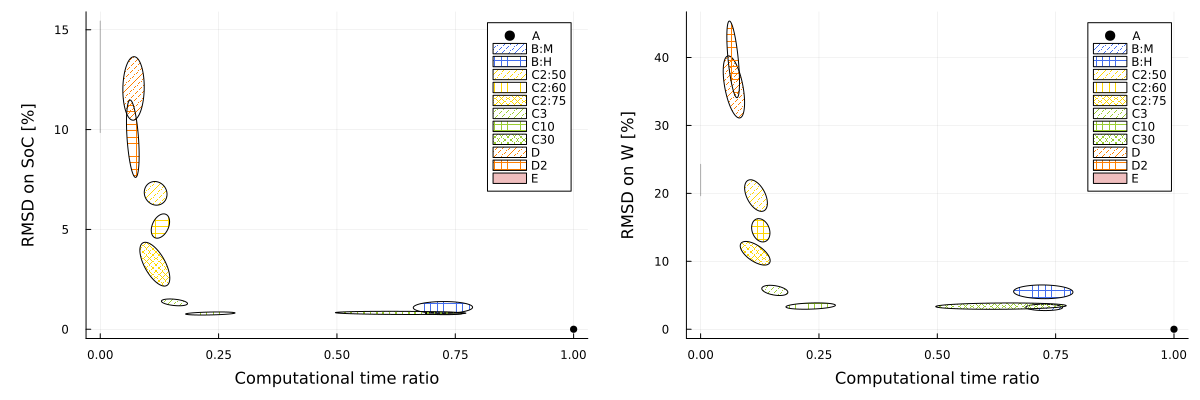}
\caption{Trade-offs between computational time and RMSD on SoC and W measured from the optimization results for all 27 price data sets. The top row shows scatter plots. The bottom row shows covariance ellipses of the models to provide performance distributions across accuracy and speed.}\label{fig:tradeoffs}
\end{figure}

Figure~\ref{fig:tradeoffs} shows the equivalent results using all 27 price data sets. The same accuracy-speed trade-off is observed. Computational times are normalized by the computational time of Model A for each price data set and run. The bottom row shows covariance ellipses of the models to provide a sense of how each model's performance is distributed across accuracy and speed. 

The Model A results have no error, by definition, and their computational time ratio is one, so they serve as the baseline. Model E shows the fastest computation but with 20\% RMSD in SoC and operations. Notably, Model E resulted in greater SoC errors than other models but less error in charging and discharging operations compared to Model D. This occurs because Model E's capability function is closer to that of Model A than Model D's is, especially in the 25\% to 75\% SoC range, as shown in Fig.~\ref{fig:charge_capability_models}. This leads to Model E's operations being more similar to Model A's and having lower operations RMSD. However, Model E's flat capability function means it is not incentivized to manage its SoC, leading to a higher SoC RMSD.

As observed in the previous section, Model B:M shows slightly better performance than Model B:H because Model A ran at full-load most of the time. The accuracy of Model C is improved by increasing the number of segments and choosing breakpoints which minimize the difference between the piecewise linear and true capability function. Model C2:75 performs better than Models C2:50 and C2:60. Models C3 and C10 achieve accuracy similar to the Model B variants, with faster computation. Models C30 and B:M have similar runtimes. Model D2 shows slightly improved accuracy over Model D, but not enough to be comparable to the Model C variants.

Lastly, we define a figure of merit (FoM) as one over the geometric distance from the origin in a three-dimensional space consisting of RMSD on SoC, RMSD on W, and the computational time ratio, as described in Eq.~\ref{eq:fom}. $RMSD_{SoC}^X$ and $RMSD_{W}^X$ are divided by 100 to match their units to computational time ratio.
\begin{equation}
\begin{split}
FoM_X = \sfrac{1}{\sqrt{{RMSD_{SoC}^{X}/100}^2 + {RMSD_{W}^{X}/100}^2 + {\text{Computational time ratio}_X}^2}}
\end{split}\label{eq:fom}
\end{equation}

Using this FoM, an ideal model would be located at the origin, indicating no error and no computation time. Model A is located at $(0,0,1)$, and any other model's performance on a sphere with radius 1 can be considered equivalently beneficial. Models closer to the origin are more beneficial than the baseline Model A, and vice versa. Note that our definition of FoM equally weighs RMSDs on SoC, W, and the computational time ratio. This is subjective and must be decided by each modeler. To give a second example, we also test a second FoM where faster computation is weighted ten times higher than accuracy.

\begin{figure}[ht]
\centering
\includegraphics[width=0.49\textwidth]{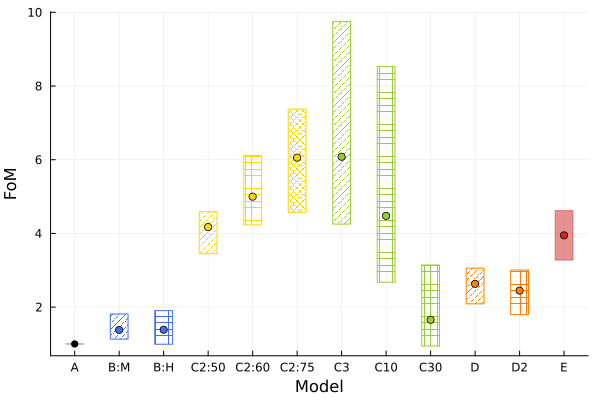}
\includegraphics[width=0.49\textwidth]{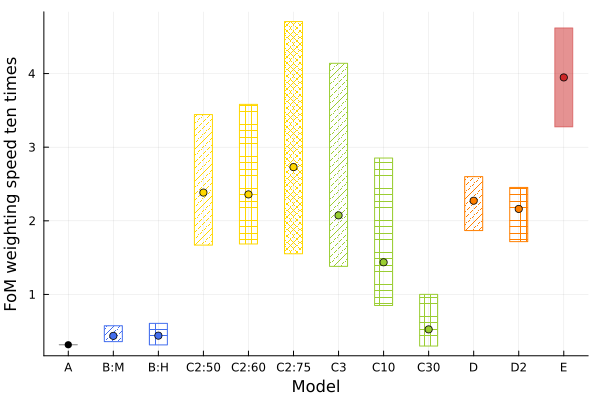}
\caption{Figure of merits (FoM) of different models. Min-max bars from 27 price data sets are shown with mean values indicated by a circle marker. The left panel shows FoM described in Eq.~\ref{eq:fom} and the right panel shows FoM that weighs computational efficiency ten times more.}\label{fig:fom_by_models}
\end{figure}

Model C3 is the best performing PTES model for most of our 27 price data sets using the equally weighted FoM, as shown in the left panel of Fig.~\ref{fig:fom_by_models}. Model C2:75 has the best worst-case performance for this FoM. Using the speed-favoring FoM, Model E has the best performance overall, excelling in both average and worst-case scenarios, while Model C2:75 has the second-best average performance and Model D has the second-best worst-case performance. The choice of FoM significantly impacts the perceived benefits of each model. However, the computational run times of the models vary more than their accuracy (100\% range vs 20\% range in Fig.~\ref{fig:tradeoffs}), so most FoM will value faster models unless accuracy is given greater weight. FoMs for each price dataset are shown in the Supplementary material.

\subsection{GenX results}\label{sec:3.3}

While the results of the price-taker case study are useful for understanding the impact of our model formulations on PTES operations and to test the formulations over many data sets, our primary interest is in how the model formulations affect investment decisions and the operation of entire grids. Therefore, we run a second case study using the GenX capacity expansion model. Most of our PTES formulations make GenX a non-linear model, greatly increasing runtimes. To ensure the model remains tractable, we use a 3-zone case study with a zero-carbon emission limit and only four technologies: PTES, utility-scale Li-ion storage, solar PV, and onshore wind. This also has the benefit of making the results more easily interpretable.

\begin{figure}[ht]
\centering
\includegraphics[width=\textwidth]{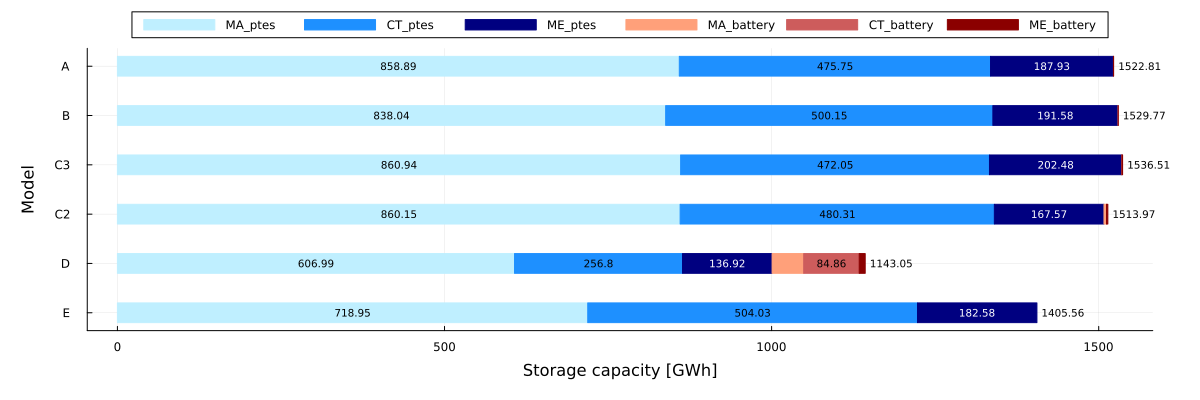}
\includegraphics[width=\textwidth]{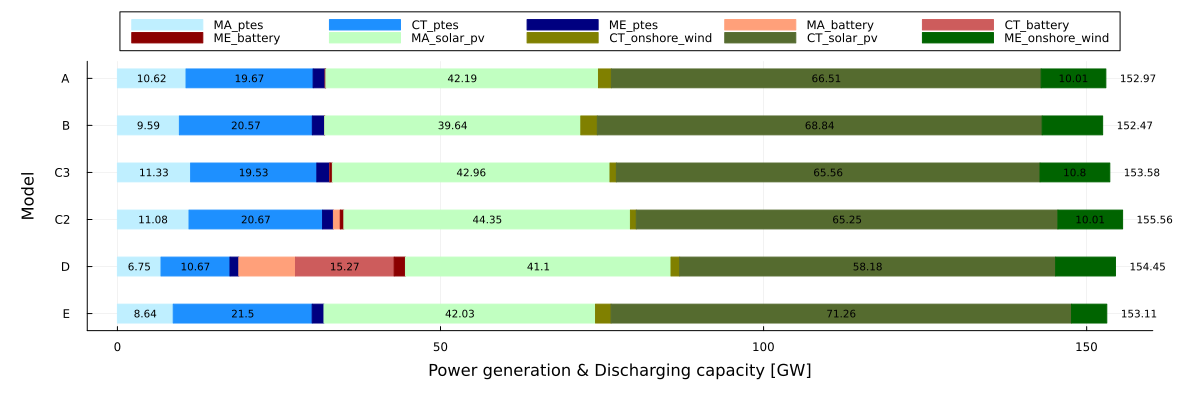}
\caption{Capacity expansion modeling results from GenX with different PTES models. The top panel displays storage capacities across three zones, while the bottom panel illustrates the power generation and discharging capacities of ten resources. Only storage capacities exceeding 50 GWh and power generation or discharging capacities above 10 GW are annotated.}
\label{fig:GenX_Capacity}
\end{figure}

\begin{figure}[ht]
\centering
\includegraphics[width=0.7\textwidth]{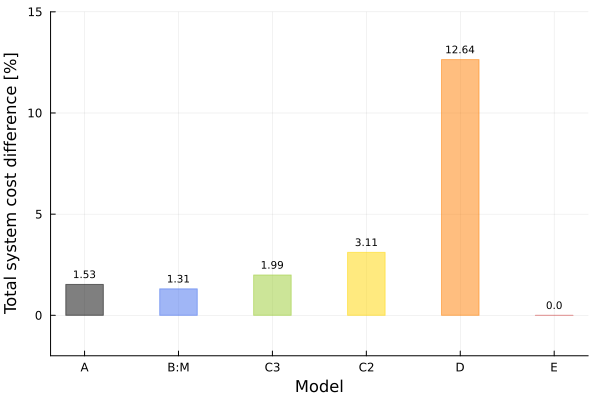}
\caption{The total system cost differences relative to the cost of Model E.}\label{fig:GenX_cTotal}
\end{figure}

Optimizing the case study using PTES Model A led to an 8\% increase in the total PTES storage capacity versus the same results using Model E. As shown in Fig.~\ref{fig:GenX_Capacity}, using Model A also led to a similar increase in the charging and discharging capacities of the PTES but did not affect the capacity of the VRE generation technologies. The additional storage was placed in the MA zone, which has the highest demand, and was used for intertemporal storage. The additional storage investments raised the overall system cost by 1.5\% (see Fig.~\ref{fig:GenX_cTotal}). The change in the system cost is less than what it might otherwise be as the grid did not have much scope to substitute different generators due to the zero-emission constraint and small pool of available generators. 

The results for the other models reinforce the claim that PTES capability functions significantly influence investment modeling results. Model E, with the best capability function, led to an optimized grid with the lowest annual costs and only PTES being used for grid energy storage, suggesting that PTES is economically more advantageous than Li-ion storage when capability functions are not considered. In contrast, using Model D, with the most pessimistic capability functions, led to an optimized grid with 35\% fewer PTES, an increased role for Li-ion storage, and 12.64\% higher grid costs. The results using Models B:M, C3, and C2 were closer to those of Model A, with Model B:M having lower costs than Model A because it is less constrained. 

\begin{figure}[ht]
\centering
\includegraphics[width=\textwidth]{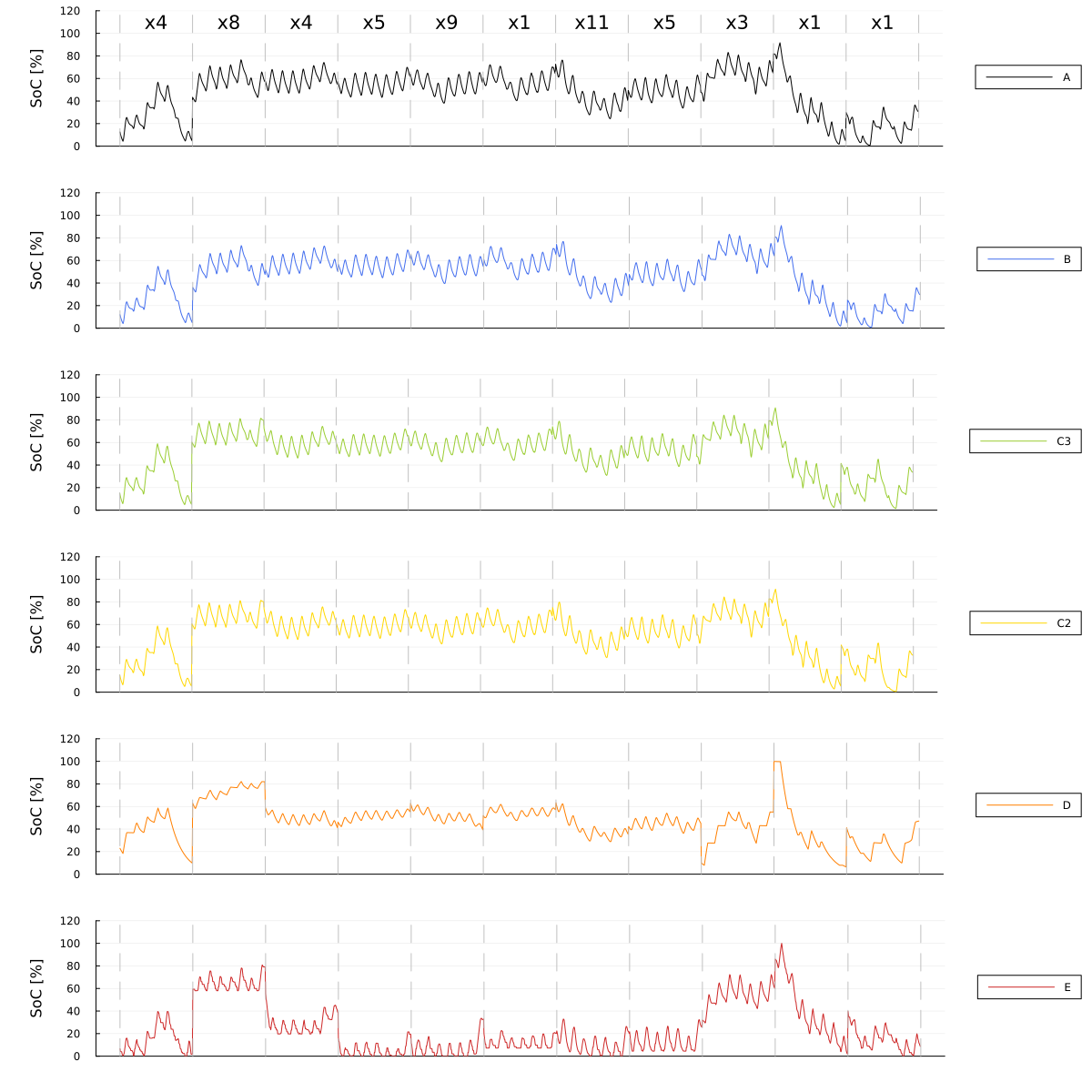}
\caption{The SoC time series of MA PTES from the optimal operation of a three-zone electricity network computed by GenX. GenX's time-domain reduction identifies 11 one-week-long representative periods. All 11 representative periods are shown horizontally and the weights for each week are annotated in the top panel.}\label{fig:GenX_SoC}
\end{figure}

\begin{figure}[ht]
\centering
\includegraphics[width=0.7\textwidth]{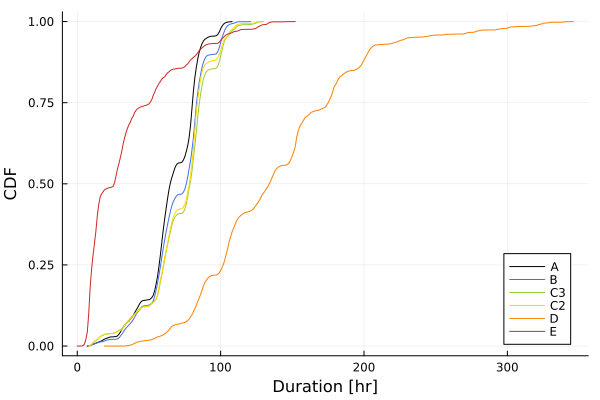}
\caption{Empirical cumulative density functions of storage durations of MA PTES for each model, recorded from the optimal hourly operations for the three-zone electricity network.}\label{fig:GenX_ecdf}
\end{figure}

The most significant difference between the Model E results and those of the other models is seen in the hourly PTES SoC and operational patterns. Fig.~\ref{fig:GenX_SoC} shows the hourly SoC for the eleven representative weeks we model, with long-duration storage constraints linking the PTES SoC between periods. Model A has an average SoC of 60\% across the entire year, with relatively slow periods of charging and discharging. Model E charges and discharges more quickly and has an average SoC of 10\% because a higher SoC is not necessary for it to charge or discharge. This lower SoC also reduces its energy leakage, reducing the total energy it must charge over the year. Fig.~\ref{fig:GenX_ecdf} shows that a Model E PTES stores 75\% of its energy for 48 hours, compared to 72 hours for a Model A, B, or C PTES. The other model formulations generally produce results similar to those of Model A. Model D exhibits the greatest differences due to its pessimistic capability function and strong incentives to keep its SoC close to 50\%.

These improvements in model accuracy come at a computational cost, as shown in Table~\ref{table:genx_cputime}. The average runtime for Models A-D is several orders of magnitude greater than that of Model E. In this simple case study these differences lead to meaningful but modest changes in the grid investments and costs. However, the large operational differences between PTES models will lead to more significant differences in more complex grid models. We recommend Models A, B, and C for capacity expansion modelling. Model D is not accurate enough to justify its runtime and Model E has a clear limitation in accurately modeling PTES operations. However, Model E remains a viable option if computational speed is a major constraint. Model C3 appears to be the best balance of computational time, investment decision accuracy and operations accuracy.

\begin{table}[ht]
\centering
\begin{tabular}{cccc}
    \hline
    Solver & Model & Iteration & Solve time [s] \\
     \hline
     \multirow{5}{8em}{\centering Ipopt\\MA86, Adaptive} & A & 601 & 1590\\
     & B:M & 560 & 1,360\\
     & C3 & 553 & 1,120\\
     & C2:75 & 534 & 1,040\\
     & D & 417 & 690\\
     \hline
     Gurobi & E & 127 & 7\\
    \hline
\end{tabular}
\caption{A summary of the computational efficiency of GenX optimization with different solvers and PTES models.}\label{table:genx_cputime}
\end{table}
\section{Conclusion}\label{sec:conclusion}

In this paper, we investigated various models of Pumped Thermal Electricity Storage (PTES) for use in price-taker and capacity expansion optimizations to assess the trade-offs between model accuracy and computational efficiency. We demonstrate that the PTES model most commonly used today (Model E in our study) produces overly optimistic capacity and operational predictions due to its disregard of SoC-dependent charging and discharging capabilities. This leads to operational errors of up to 50\% compared to our most physically accurate model (Model A). However, Model E is at approximately 200x faster to optimize. Our intermediate Model B and C variants simplify the capability function of Model A while retaining reasonable physical accuracy, present a balanced trade-off between computational complexity and accuracy. In particular, the Model C variants produce price-taker and capacity expansion results with less the 5\% error and runtimes which are 33-80\% less than when using Model A.

In future work, we will develop strategies to accelerate the Model C variants. We believe it is possible to adapt Model C to a linear formulation using a bi-level approach or by slightly restricting the design space of the system. This will greatly close the gap between it and the Model E runtime without compromising the more accurate PTES operations and grid investments. Once this is achieved, we will add further details to the PTES model, particularly the representation of energy leakage, and produce capability functions for other storage technologies. Finally, we will incorporate all of these improvements into GenX and demonstrate their impact in a larger case study.

In summary, our findings suggest that while physically accurate models are ideal for detailed operational studies, simplified models like Model C variants offer a practical alternative for larger-scale investment modeling and decision-making, maintaining a balance between accuracy and computational performance.

\clearpage

\appendix
\section{PTES configuration}\label{app1}
In our case study, we used a pumped thermal electricity storage (PTES) system with two packed beds, combining a mathematically optimized model developed by \citet{frate2022techno} and a commissioned system demonstrated by \citet{ameen2023demonstration}. Thermodynamic parameters from \citet{frate2022techno} were optimized to maximize round-trip efficiency, and Argon was selected as the working fluid. The specifications of the thermal stores were taken from \citet{ameen2023demonstration}. Magnetite pebbles (86\% Fe$_3$O$_4$ and 14\% SiO$_2$) were used as the storage medium in the stores. Finally, we determined the total volume of storage, as well as the nominal charging and discharging power, for the price-taker case study. All values are summarized in Table~\ref{table:PTES_configuration}.

% \begin{tabular}{cc}
%   \hline
%   Parameter & Value\\
%   \hline
%   $T_{ch,comp,out}$, $T_{dis,exp,in}$ & 590 $^{\circ}$C\\
%   $T_{ch,exp,out}$, $T_{dis,comp,in}$ & -100 $^{\circ}$C\\
%   $T_{ch,comp,in}$, $T_{dis,exp,out}$ & 166 $^{\circ}$C\\
%   $T_{ch,exp,in}$, $T_{dis,comp,out2}$ & 25 $^{\circ}$C\\
%   $T_{dis,comp,out1}$ & 124 $^{\circ}$C\\
%   $c_{p,Ar}$ & 0.5203 kJ/kg/K\\
%   $\alpha_{ch}$ & 1.89\\
%   $\alpha_{dis}$ & 2.83\\
%   \hline
%   Average density, $\rho_{Magnetite}$ & 4800 kg/m$^3$\\
%   Void fraction, $\epsilon$ & 0.425\\
%   $c_{p,Magnetite}$ & 0.848 kJ/kg/K\\
%   \hline
%   Total volume, $V$ & 30 m$^3$\\
%   $\overline{W}_{ch}$ & 250 kW\\
%   $\overline{W}_{dis}$ & 160 kW\\
%   $C$ & 11,021 kWh\\
%   \hline
% \end{tabular}
% \caption{Configuration parameters of the PTES system used in the case study.}\label{table:PTES_configuration}

\begin{center}
  \begin{tabular}{ll}
    \hline
    \textbf{Parameter} & \textbf{Value}\\
    \hline
    $T_{ch,comp,out}$, $T_{dis,exp,in}$ & 590 $^{\circ}$C\\
    $T_{ch,exp,out}$, $T_{dis,comp,in}$ & -100 $^{\circ}$C\\
    $T_{ch,comp,in}$, $T_{dis,exp,out}$ & 166 $^{\circ}$C\\
    $T_{ch,exp,in}$, $T_{dis,comp,out2}$ & 25 $^{\circ}$C\\
    $T_{dis,comp,out1}$ & 124 $^{\circ}$C\\
    $c_{p,Ar}$ & 0.5203 kJ/kg/K\\
    $\alpha_{ch}$ & 1.89\\
    $\alpha_{dis}$ & 2.83\\
    \hline
    Average density, $\rho_{Magnetite}$ & 4800 kg/m$^3$\\
    Void fraction, $\epsilon$ & 0.425\\
    $c_{p,Magnetite}$ & 0.848 kJ/kg/K\\
    \hline
    Total volume, $V$ & 30 m$^3$\\
    $\overline{W}_{ch}$ & 250 kW\\
    $\overline{W}_{dis}$ & 160 kW\\
    $C$ & 11,021 kWh\\
    \hline
  \end{tabular}
  \captionof{table}{Configuration parameters of the PTES system used in the case study.}\label{table:PTES_configuration}
\end{center}

\clearpage

\section{Charging and discharging capabilities function fitting results}\label{app2}
\begin{figure}[h]
\centering
\includegraphics[width=\textwidth]{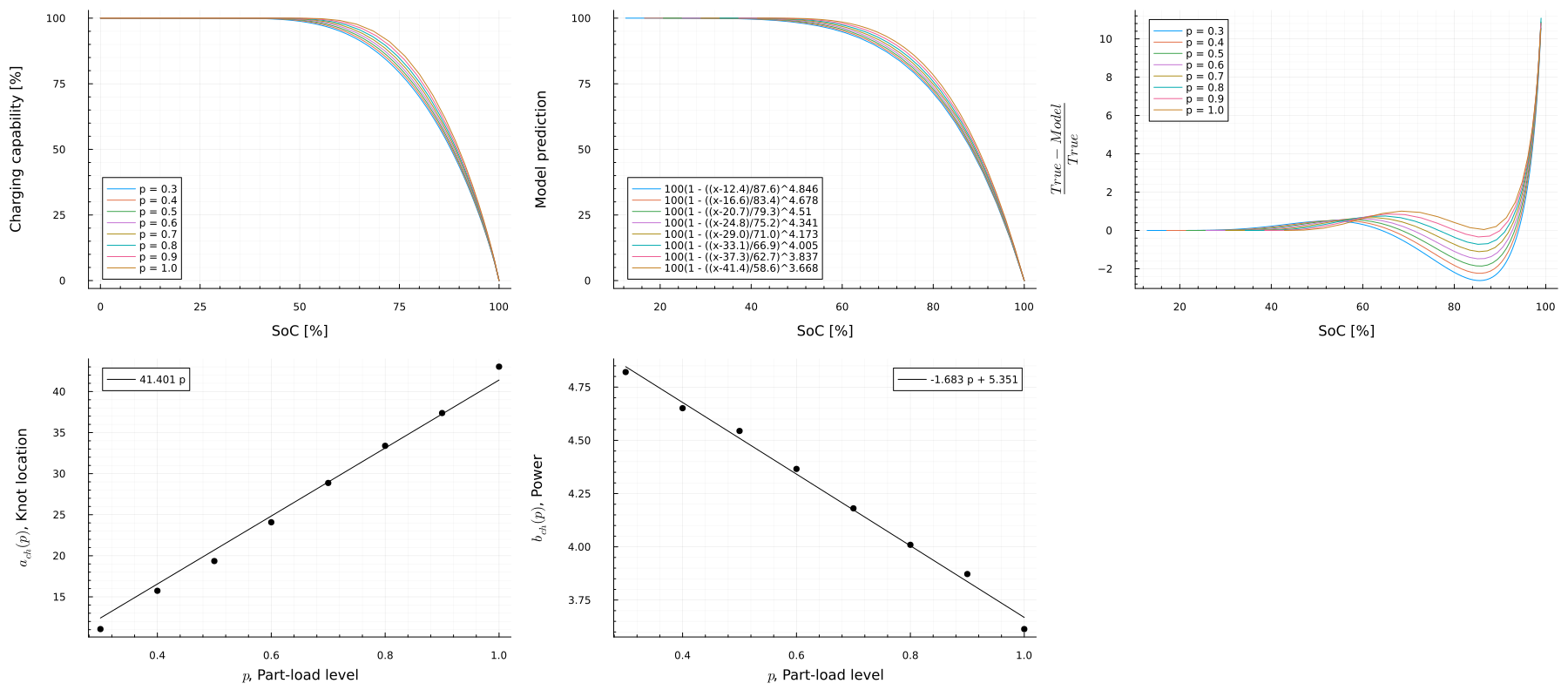}
\caption{Function fitting result for charging capability function of Model A.}\label{fig:charging}
\end{figure}

\begin{figure}[h]
\centering
\includegraphics[width=\textwidth]{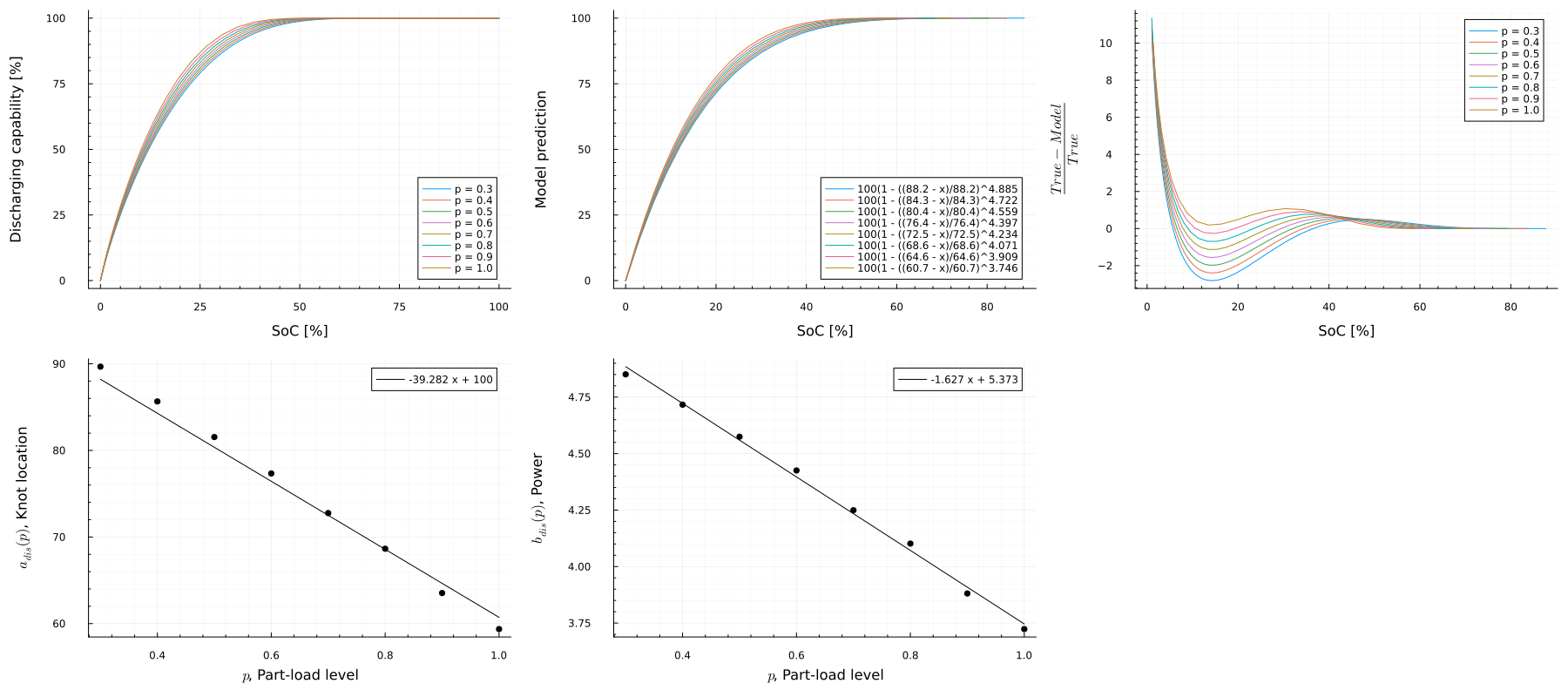}
\caption{Function fitting result for discharging capability function of Model A.}\label{fig:discharging}
\end{figure}

In both Figs.~\ref{fig:charging} and \ref{fig:discharging}, the left panel of the top row shows the typical capability curves, and the center panel shows the fitted functions. The right panel of the top row illustrates the difference between the true and fitted curves. Finally, the bottom panel shows the linear relationships between the part-load level and the two parameters: knot location and power.

\clearpage

\section{Hourly day-ahead LMP data}\label{app3}
\begin{figure}[h]
\centering
\includegraphics[width=0.32\textwidth]{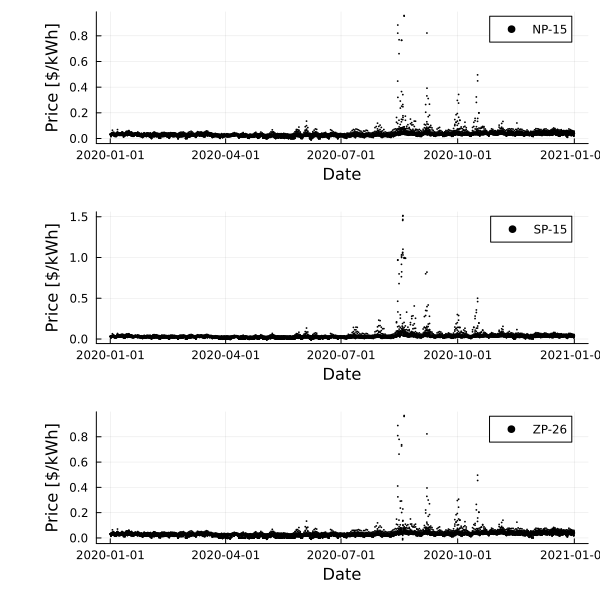}
\includegraphics[width=0.32\textwidth]{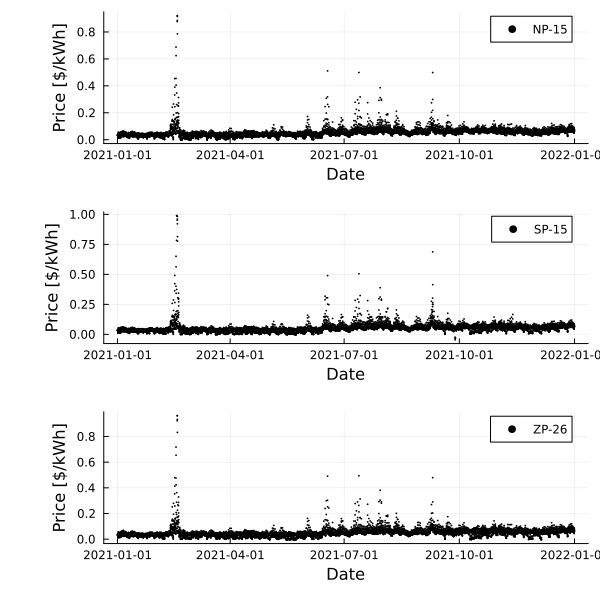}
\includegraphics[width=0.32\textwidth]{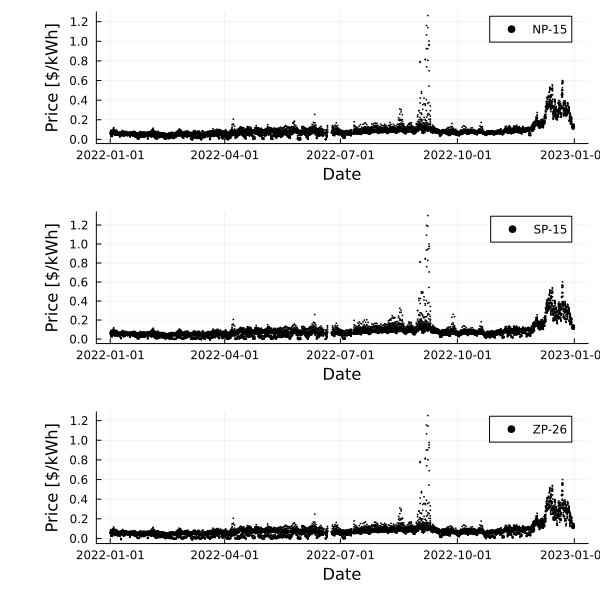}
\caption{Hourly day-ahead LMP data for CAISO in 2020, 2021, and 2022. Note that some data is missing for June 2022.}\label{fig:caiso}
\end{figure}

\begin{figure}[h]
\centering
\includegraphics[width=0.32\textwidth]{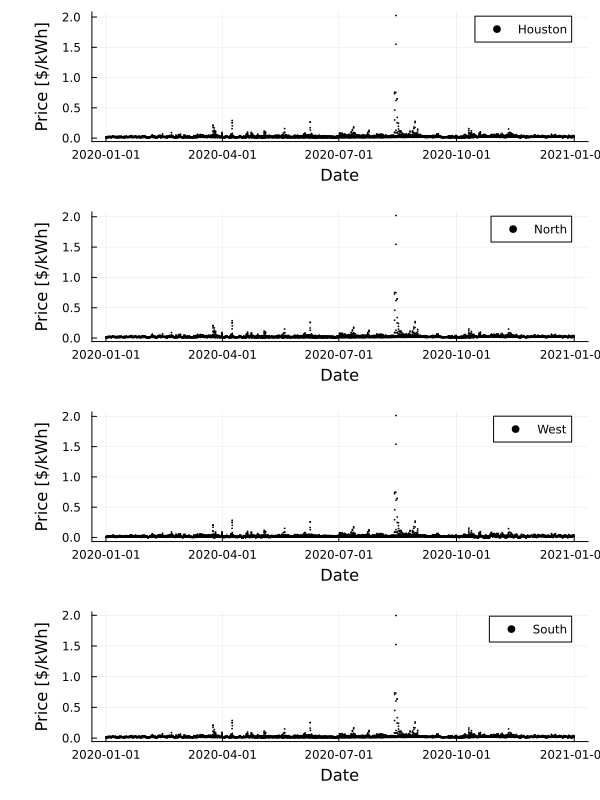}
\includegraphics[width=0.32\textwidth]{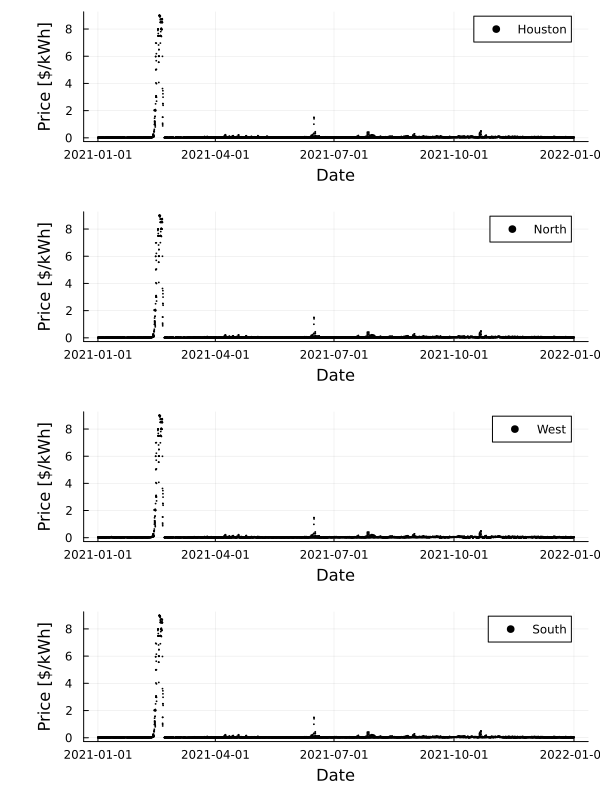}
\includegraphics[width=0.32\textwidth]{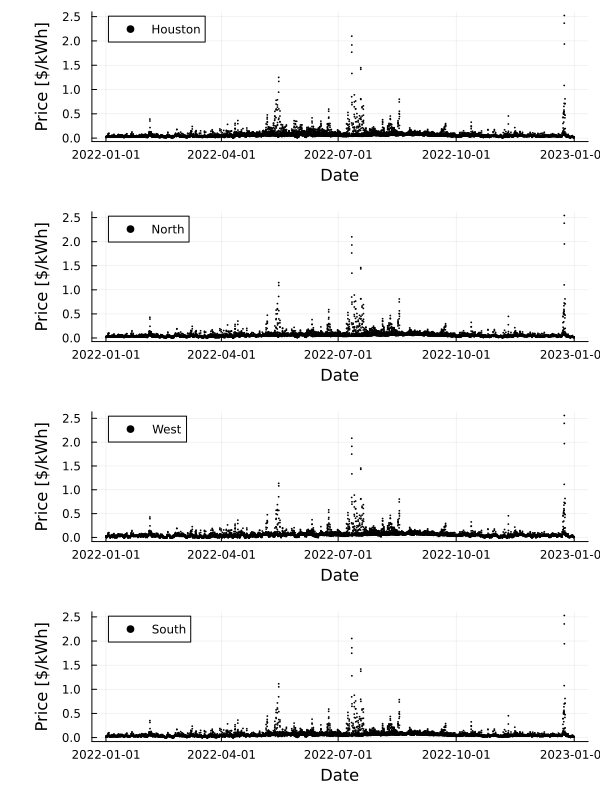}
\caption{Hourly day-ahead LMP data for ERCOT in 2020, 2021, and 2022. Note the extreme price spikes during the statewide power crisis caused by the winter storm in 2021.}\label{fig:ercot}
\end{figure}

\begin{figure}[h]
\centering
\includegraphics[width=0.32\textwidth]{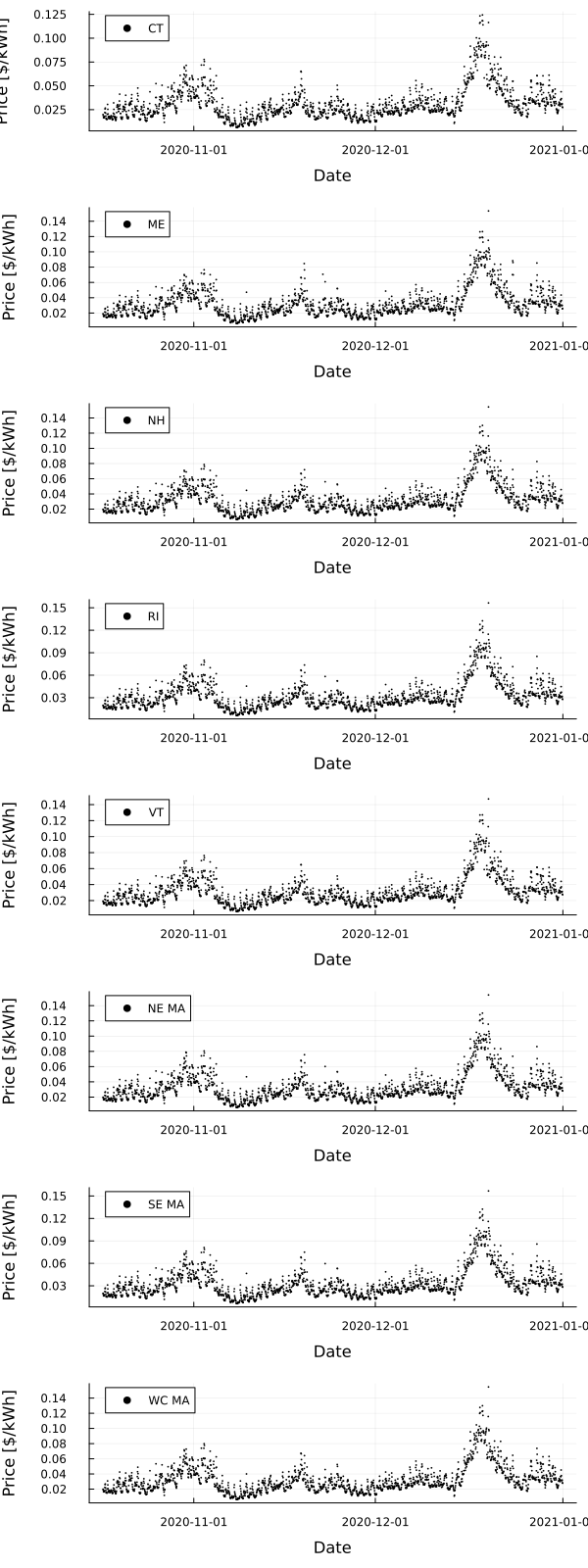}
\includegraphics[width=0.32\textwidth]{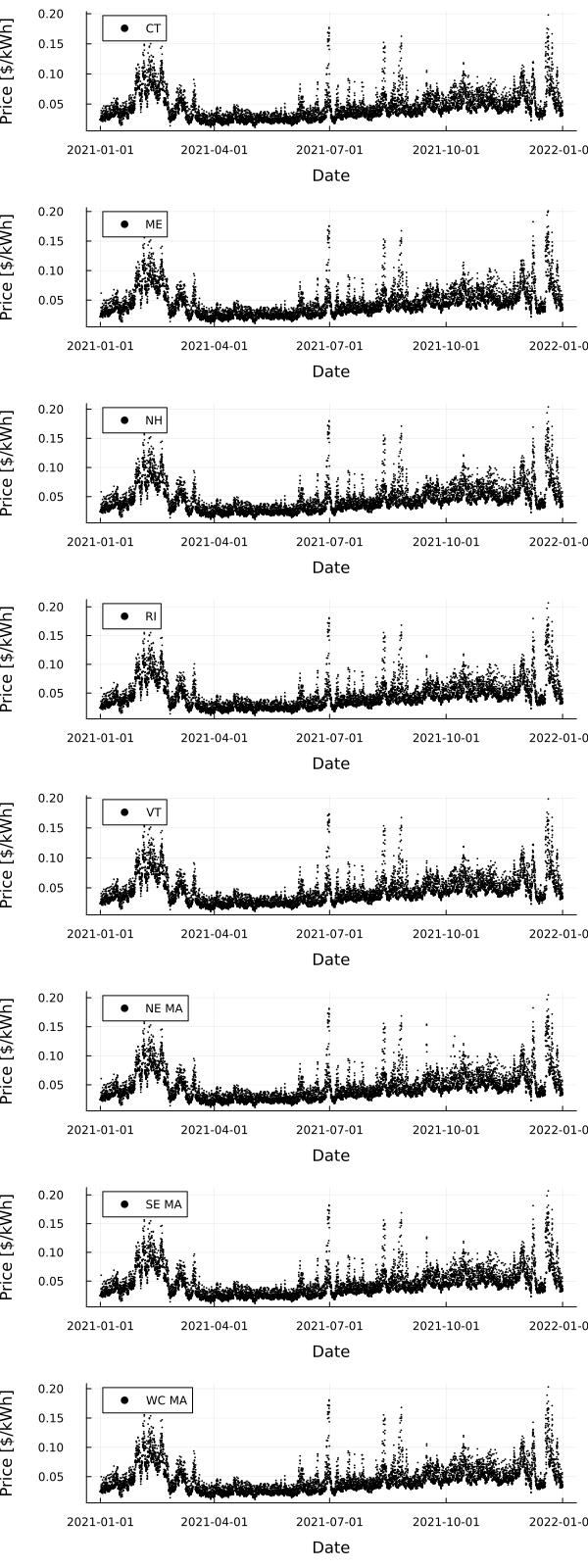}
\includegraphics[width=0.32\textwidth]{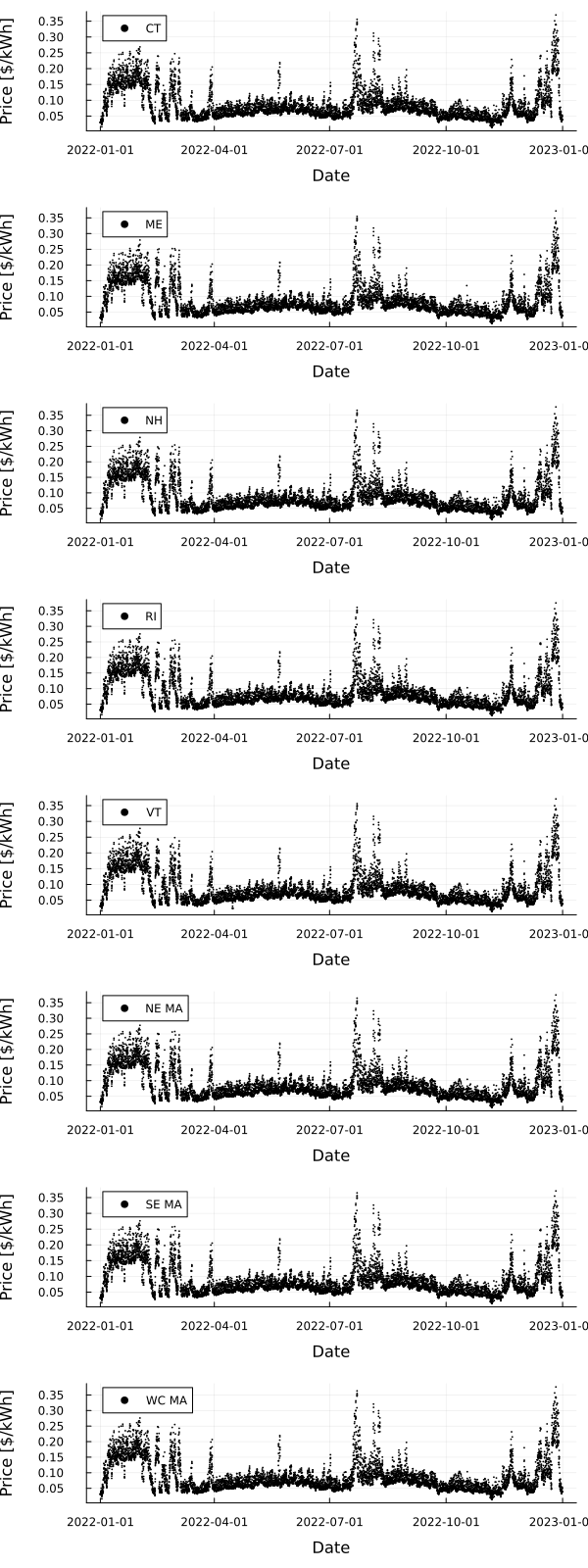}
\caption{Hourly day-ahead LMP data for ISONE in 2020, 2021, and 2022.}\label{fig:isone}
\end{figure}

\begin{figure}[h]
\centering
\includegraphics[width=0.32\textwidth]{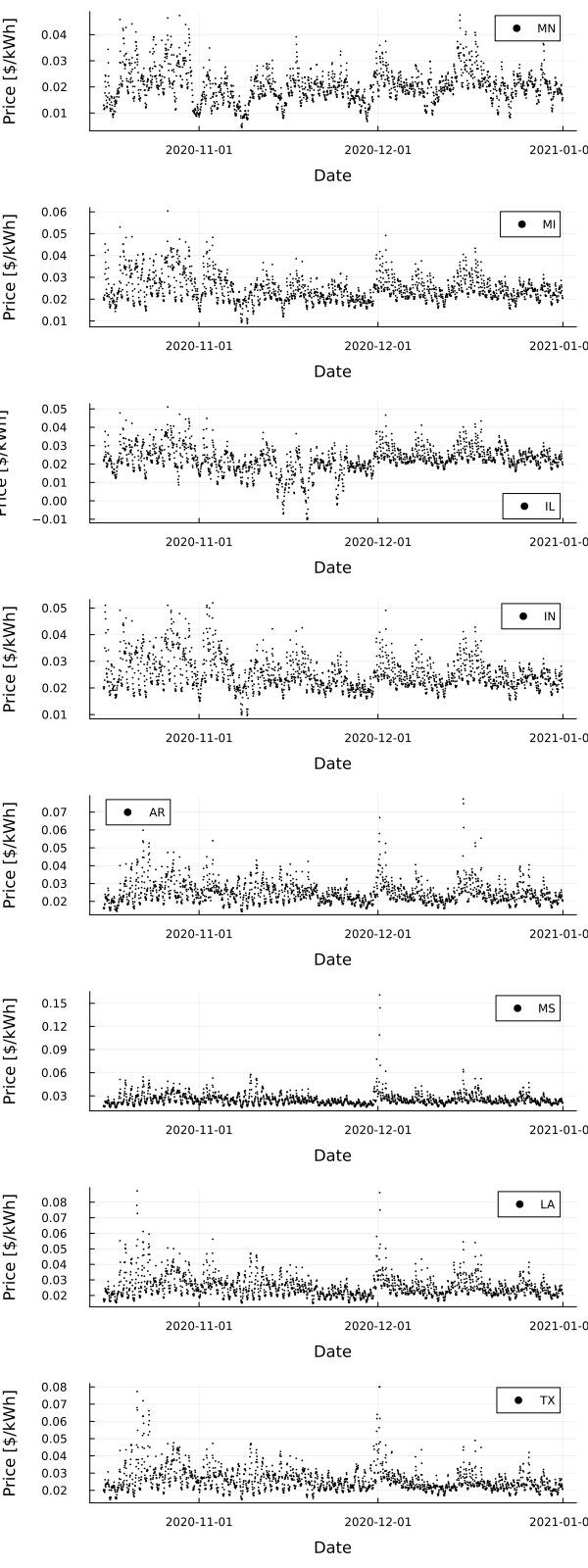}
\includegraphics[width=0.32\textwidth]{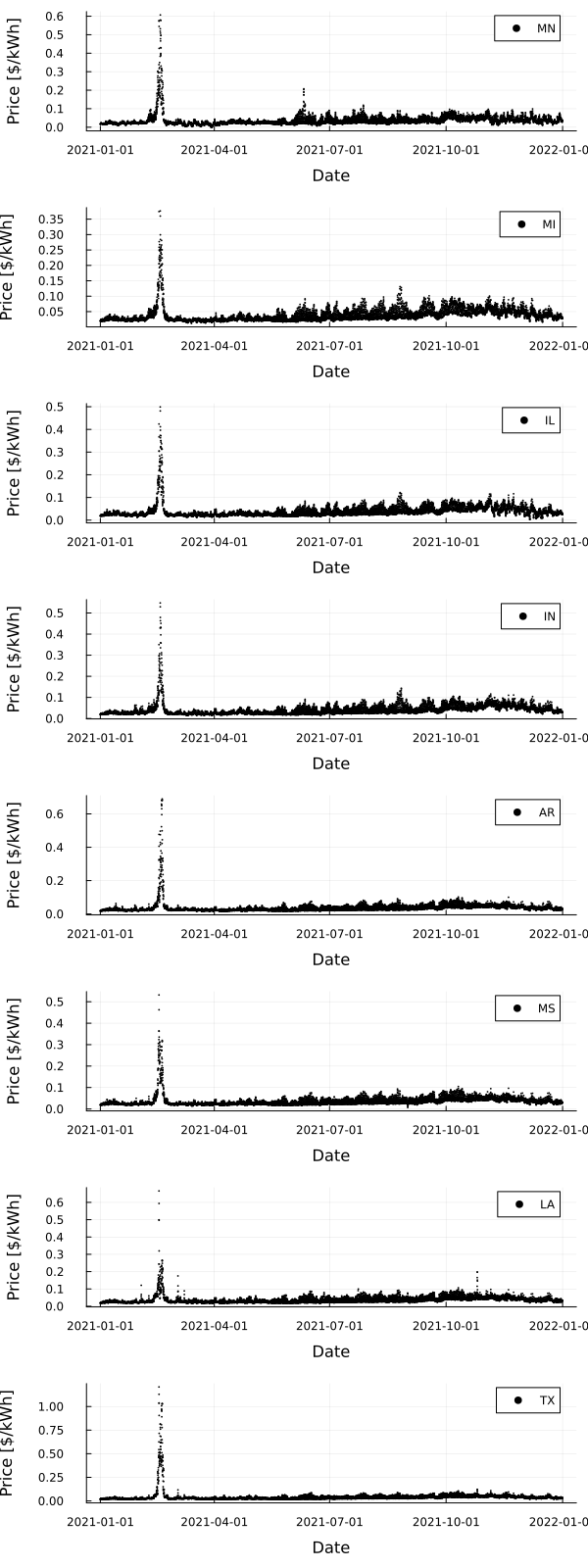}
\includegraphics[width=0.32\textwidth]{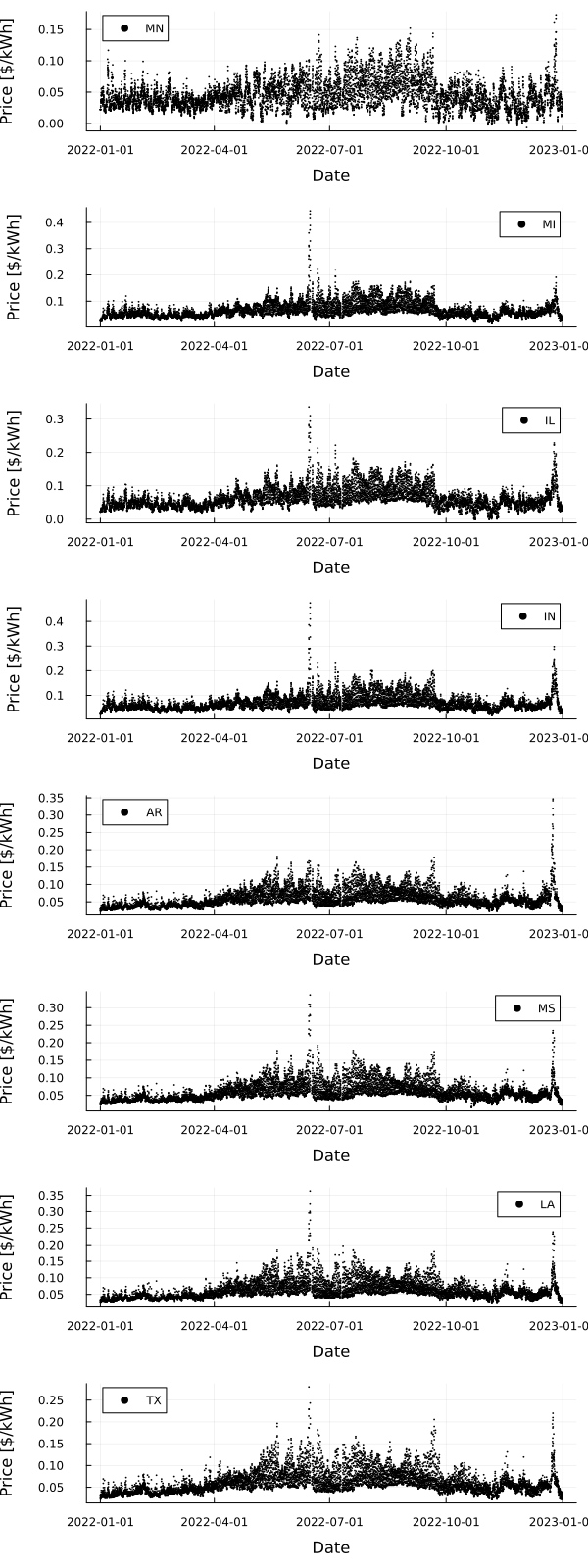}
\caption{Hourly day-ahead LMP data for MISO in 2020, 2021, and 2022.}\label{fig:miso}
\end{figure}

\begin{figure}[h]
\centering
\includegraphics[width=0.32\textwidth]{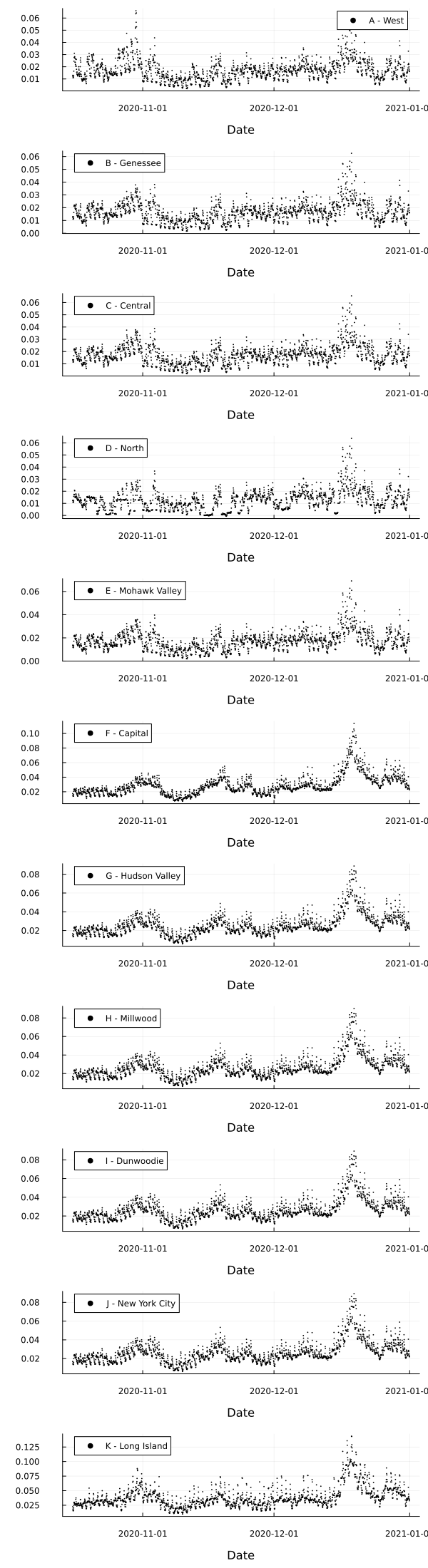}
\includegraphics[width=0.32\textwidth]{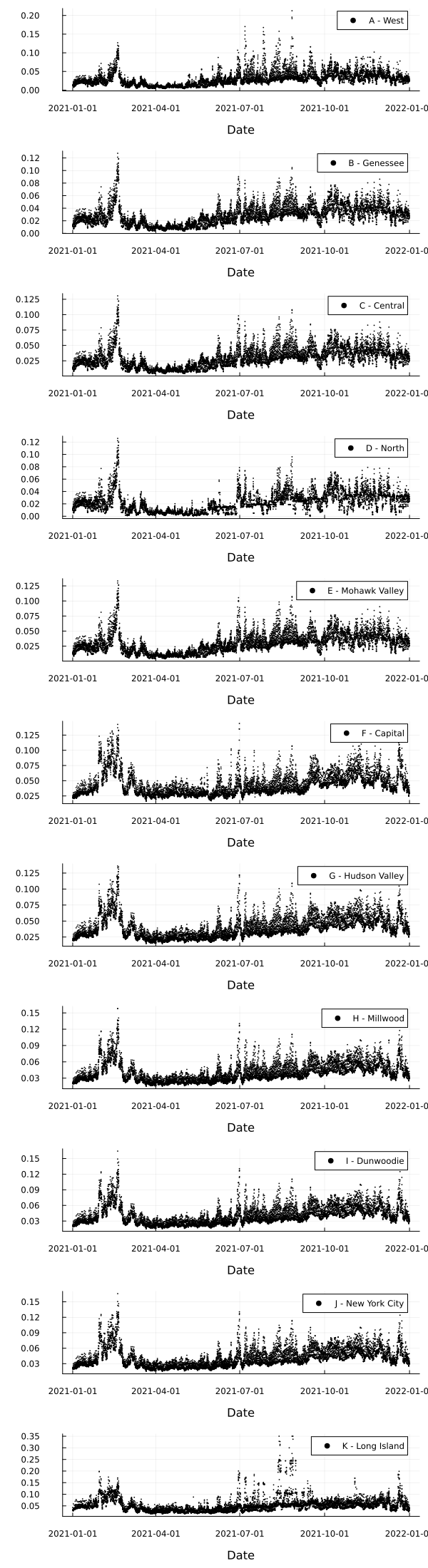}
\includegraphics[width=0.32\textwidth]{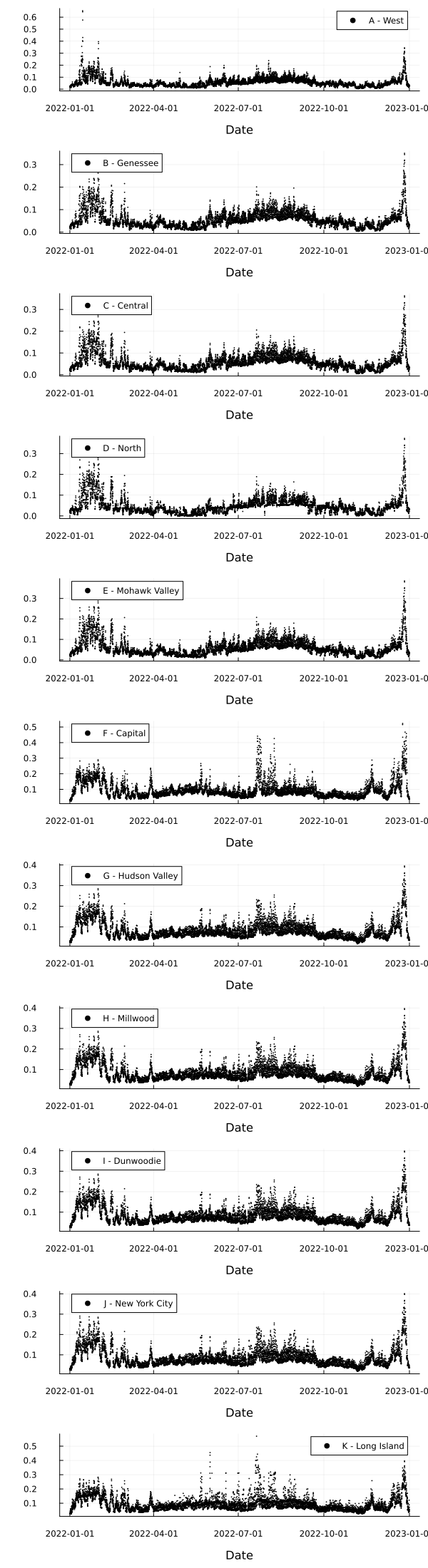}
\caption{Hourly day-ahead LMP data for NYISO in 2020, 2021, and 2022.}\label{fig:nyiso}
\end{figure}

\begin{figure}[h]
\centering
\includegraphics[width=0.32\textwidth]{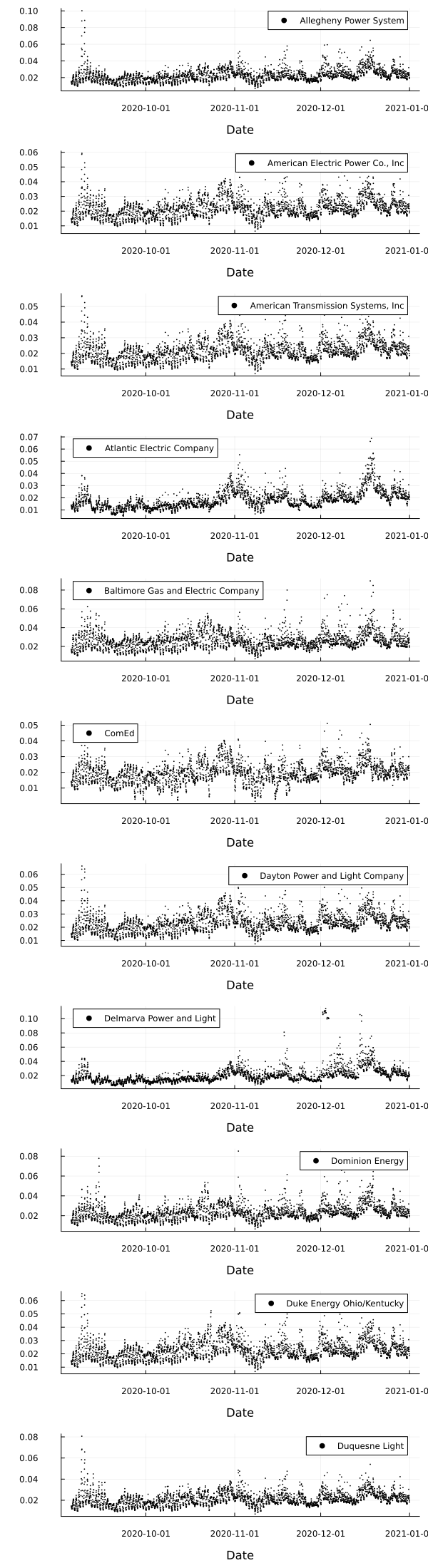}
\includegraphics[width=0.32\textwidth]{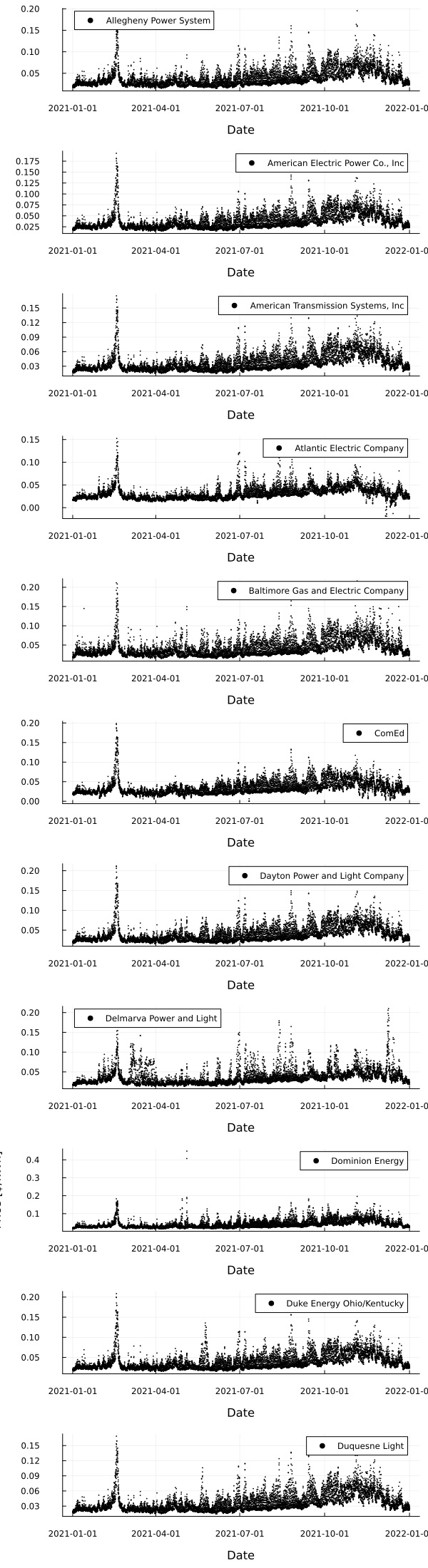}
\includegraphics[width=0.32\textwidth]{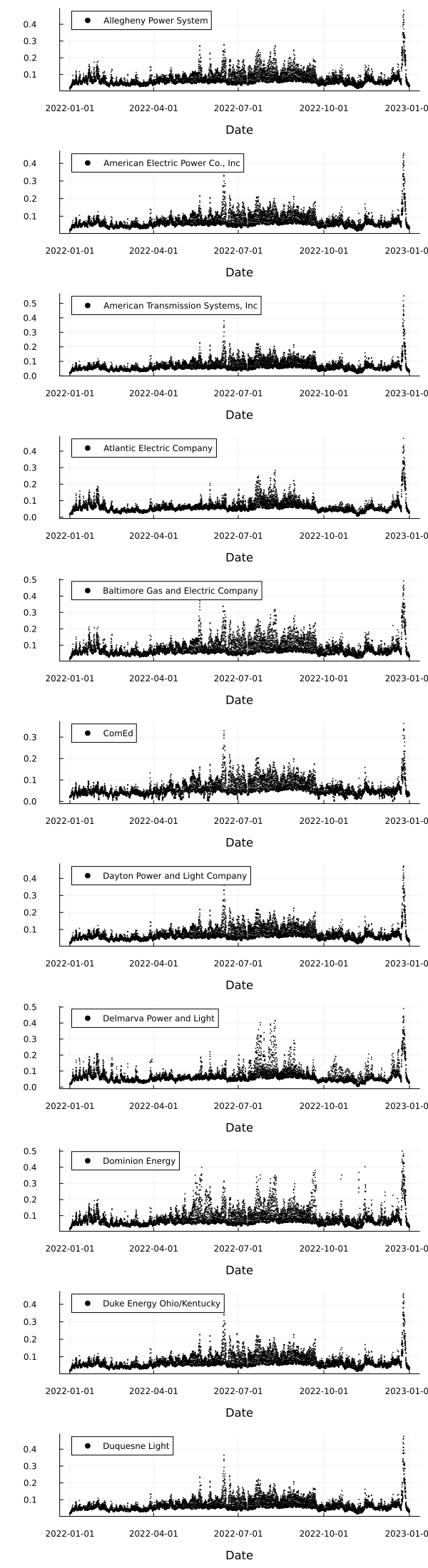}
\caption{Hourly day-ahead LMP data for the first eleven zones of PJM in 2020, 2021, and 2022.}\label{fig:pjm1}
\end{figure}

\begin{figure}[h]
\centering
\includegraphics[width=0.32\textwidth]{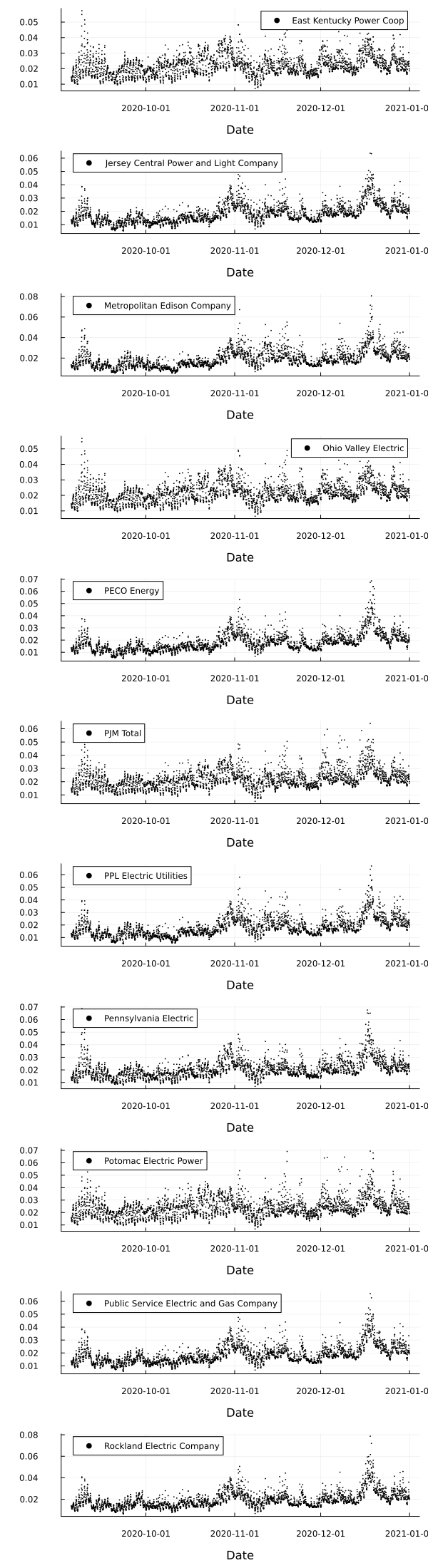}
\includegraphics[width=0.32\textwidth]{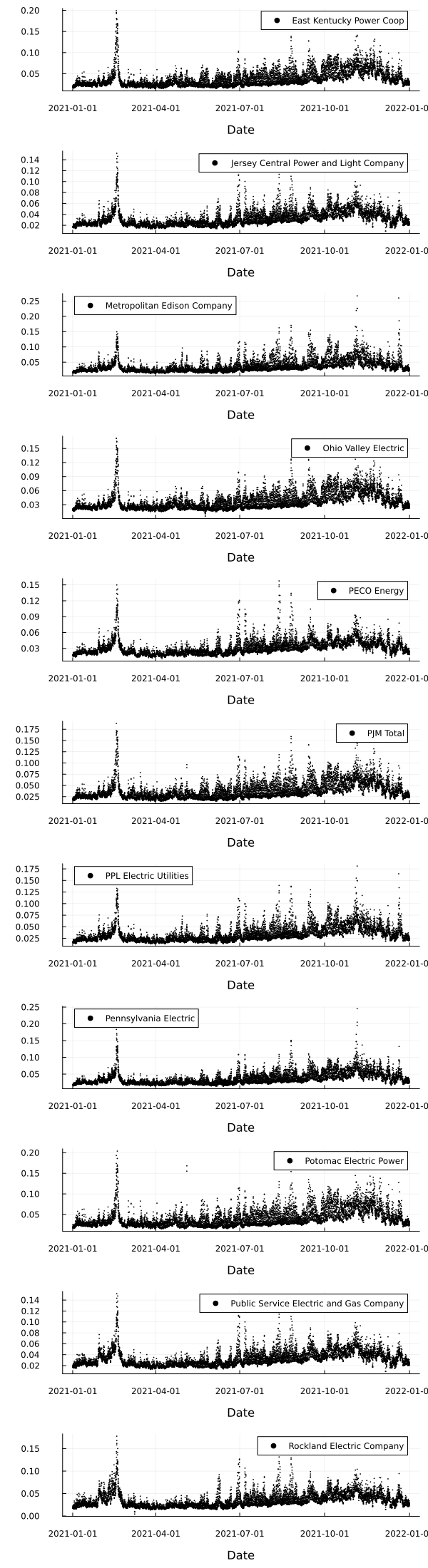}
\includegraphics[width=0.32\textwidth]{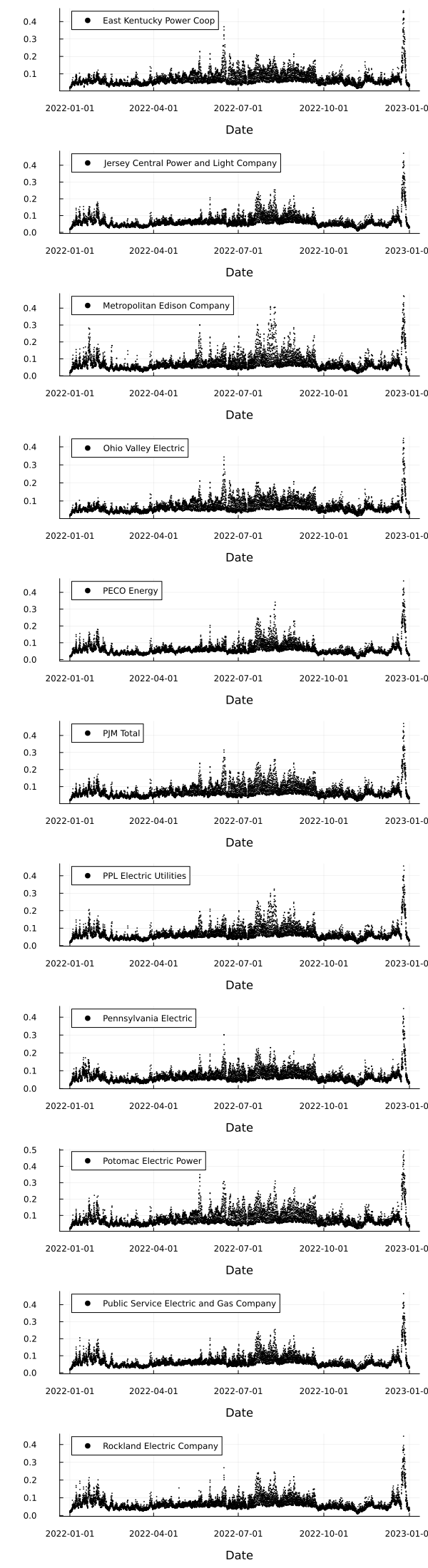}
\caption{Hourly day-ahead LMP data for the second eleven zones of PJM in 2020, 2021, and 2022.}\label{fig:pjm2}
\end{figure}

\begin{figure}[h]
\centering
\includegraphics[width=0.32\textwidth]{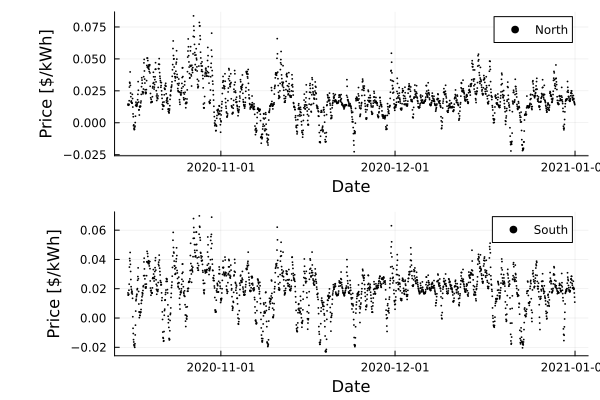}
\includegraphics[width=0.32\textwidth]{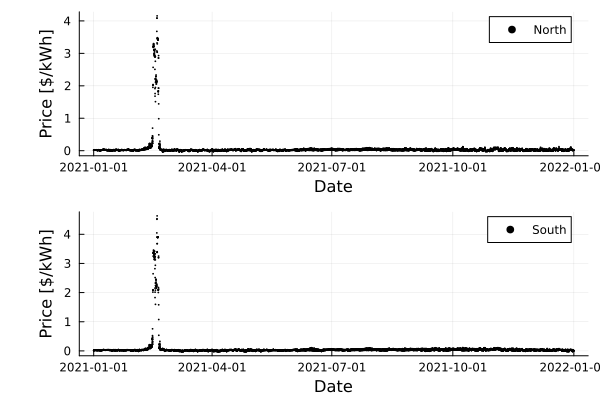}
\includegraphics[width=0.32\textwidth]{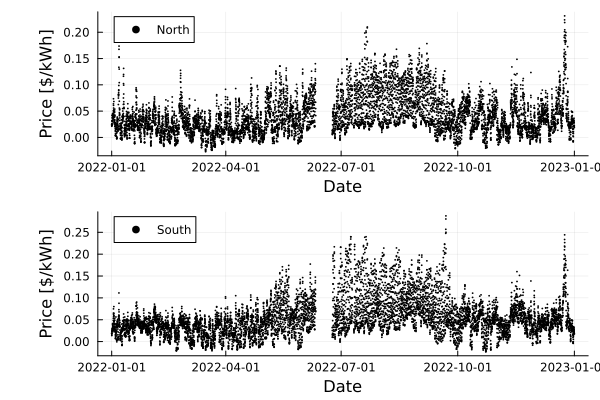}
\caption{Hourly day-ahead LMP data for SPP in 2020, 2021, and 2022. Note that some data is missing for June 2022.}\label{fig:spp}
\end{figure}

The U.S. Energy Information Administration provides wholesale electricity market data for the seven Regional Transmission Organizations (RTOs) and Independent System Operators (ISOs)~\cite{eia}. We collected hourly day-ahead Locational Marginal Prices (LMP) data from all seven RTOs and ISOs for the years 2020 to 2022, as shown in Figs.~\ref{fig:caiso}-\ref{fig:spp}. Note that zone-to-zone variations are relatively minor compared to year-to-year and ISO-to-ISO variations.

\clearpage

\section{Model comparison using a single LMP time series}\label{app4}
Optimal hourly operations derived from all models for the ERCOT North hub in 2022 are shown in this section. In all figures, the top panel displays the charging and discharging power over the first 100 hours. The second panel shows the LMP time series, while the third panel presents the State of Charge (SoC) time series. The bottom panel illustrates the SoC difference between Model A and each of the other models.

\begin{figure}[h]
\centering
\includegraphics[width=0.32\textwidth]{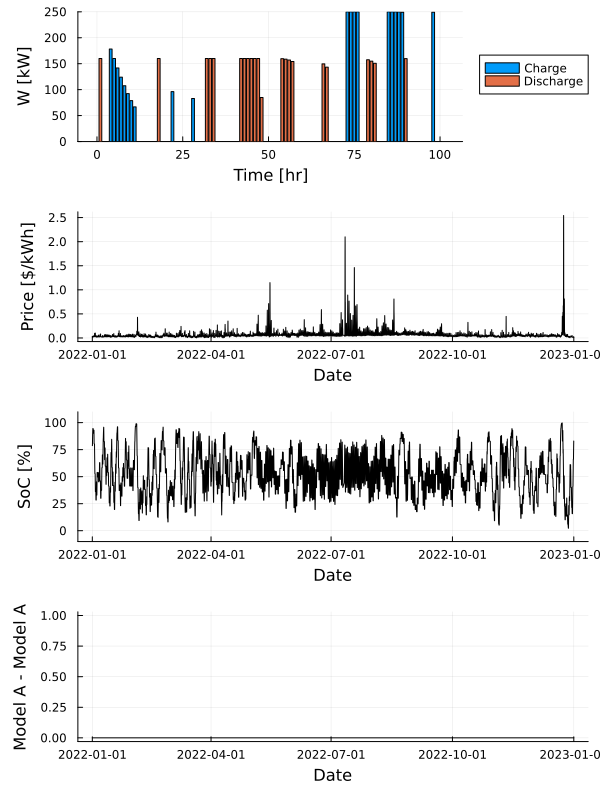}
\includegraphics[width=0.32\textwidth]{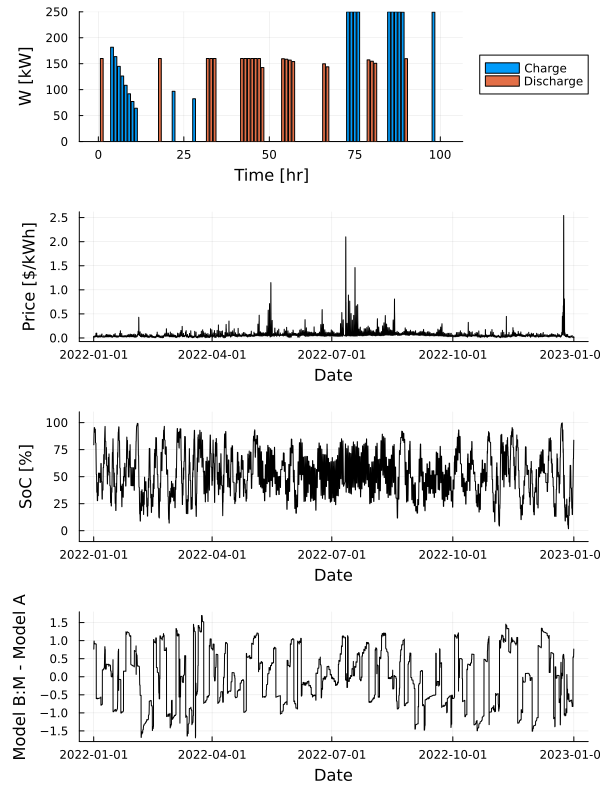}
\includegraphics[width=0.32\textwidth]{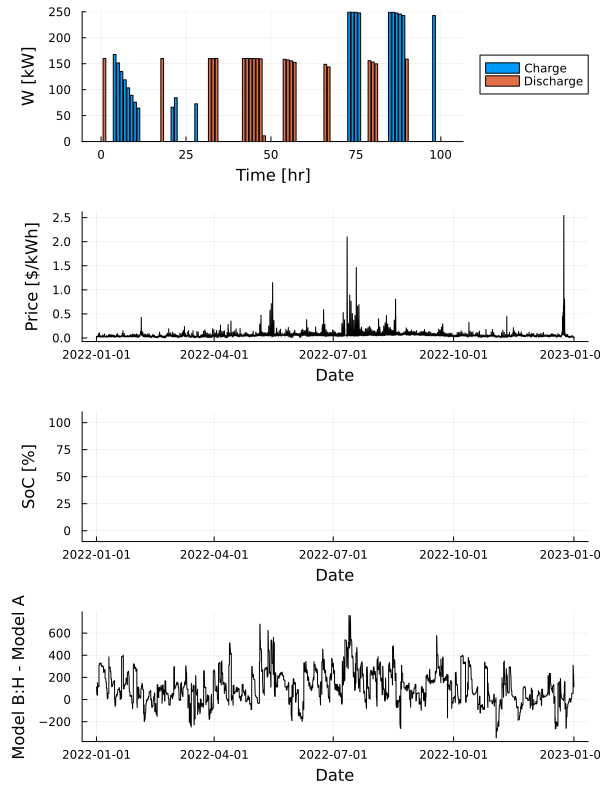}
\caption{Optimal hourly operations derived from Models A, B:M, and B:H.}
\end{figure}

\begin{figure}[h]
\centering
\includegraphics[width=0.32\textwidth]{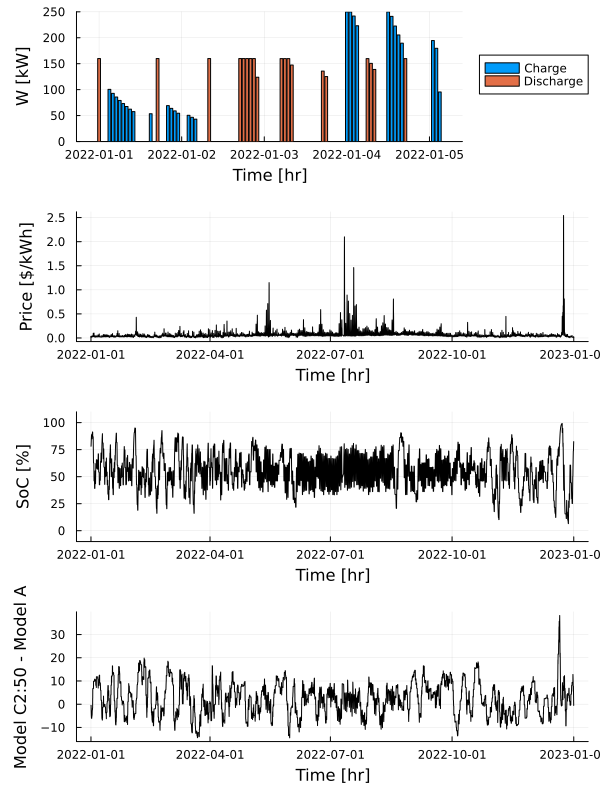}
\includegraphics[width=0.32\textwidth]{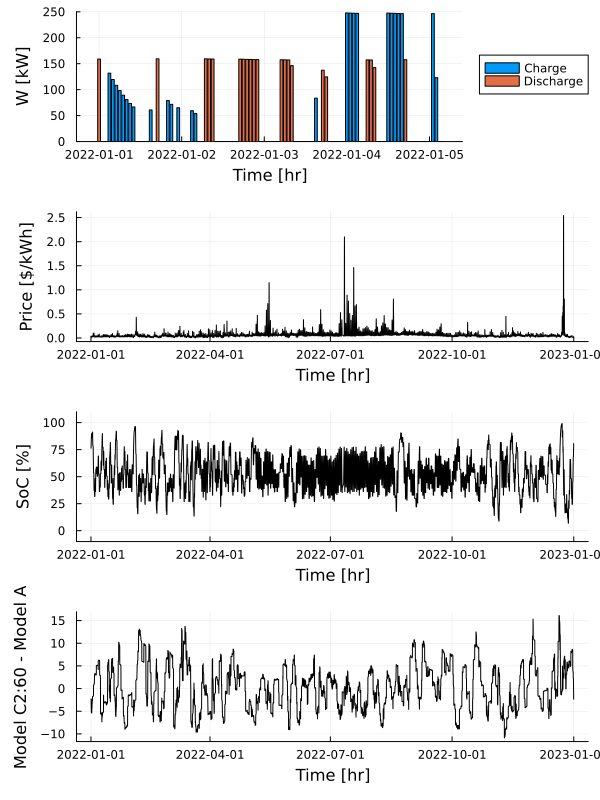}
\includegraphics[width=0.32\textwidth]{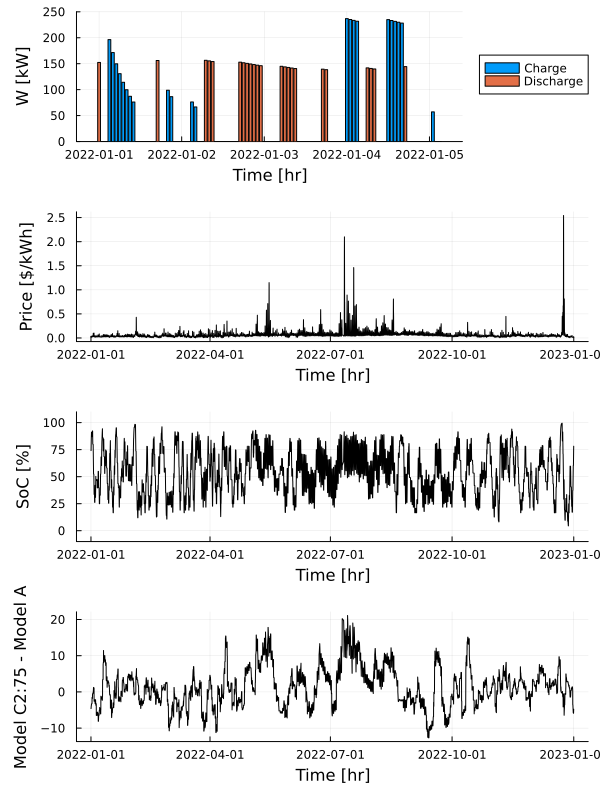}
\caption{Optimal hourly operations derived from Models C2:50, C2:60, and C2:75.}
\end{figure}

\begin{figure}[h]
\centering
\includegraphics[width=0.32\textwidth]{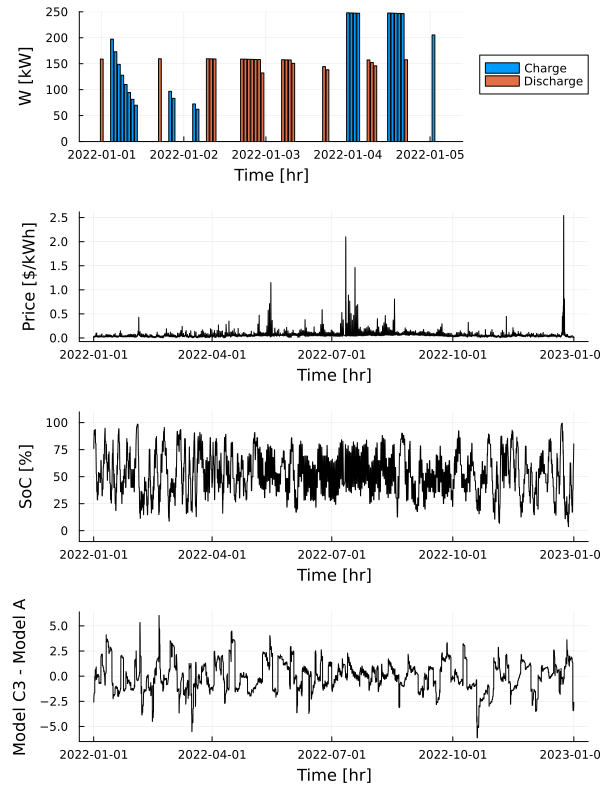}
\includegraphics[width=0.32\textwidth]{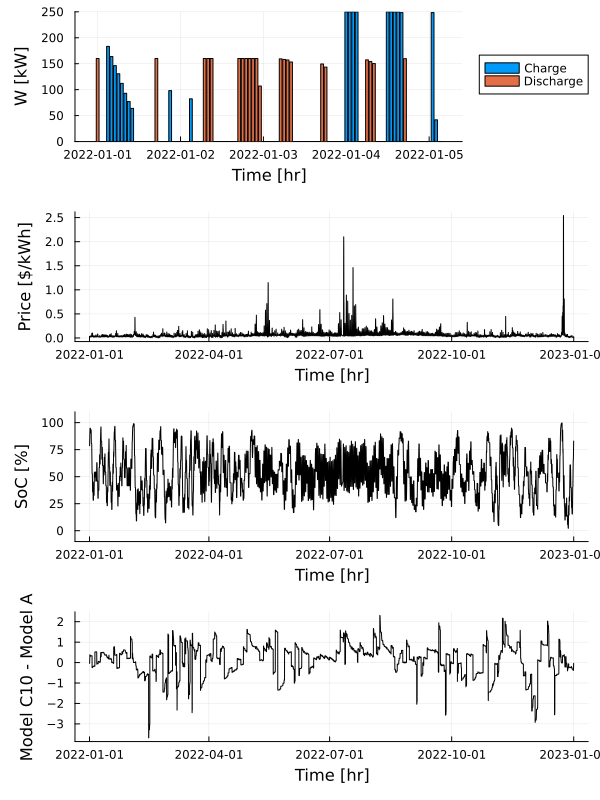}
\includegraphics[width=0.32\textwidth]{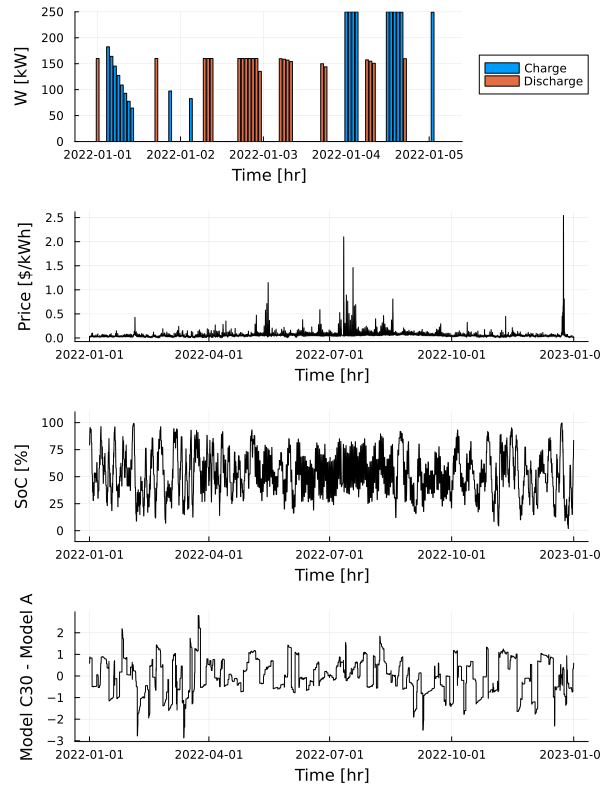}
\caption{Optimal hourly operations derived from Models C3, C10, and C30.}
\end{figure}

\begin{figure}[h]
\centering
\includegraphics[width=0.32\textwidth]{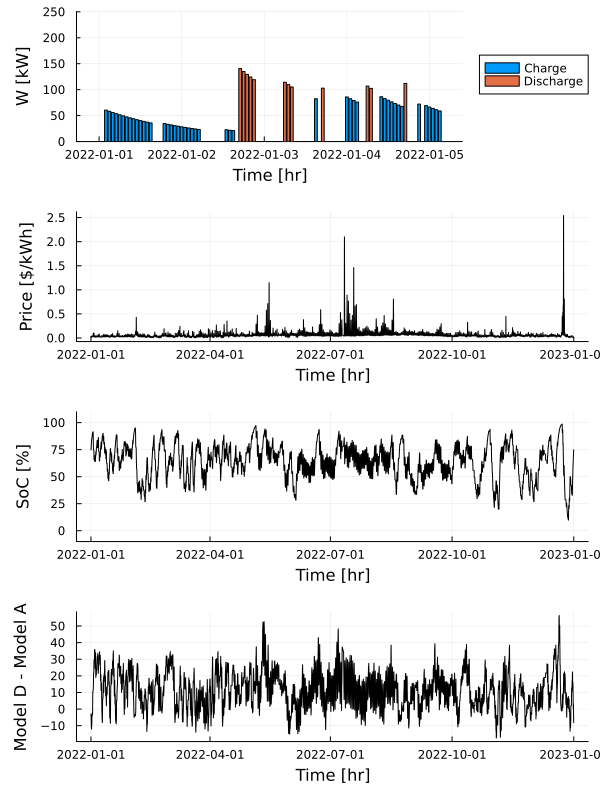}
\includegraphics[width=0.32\textwidth]{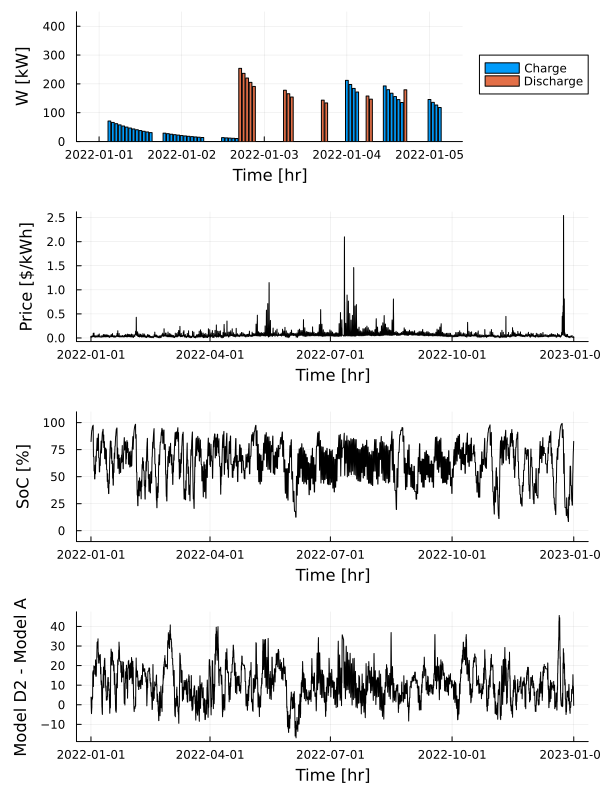}
\includegraphics[width=0.32\textwidth]{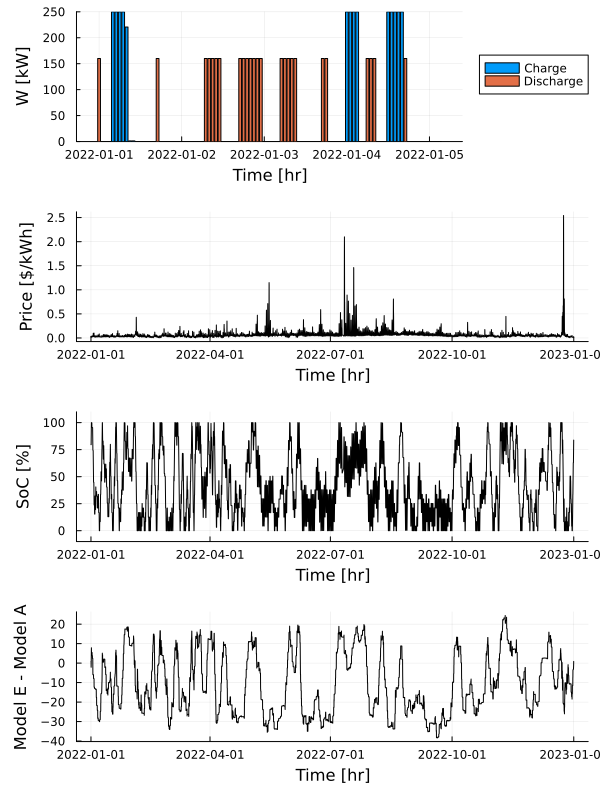}
\caption{Optimal hourly operations derived from Models D, D2, and E.}
\end{figure}

\clearpage

\section{Energy storage duration distributions}\label{app5}
We monitored the energy input (charging) to the PTES and the energy output (discharging) from the PTES in a first-come, first-served manner to track the storage duration of every watt of energy. The empirical cumulative density functions (ECDFs) of all 27 price data sets are shown in this section.

\begin{figure}[h]
\centering
\includegraphics[width=0.32\textwidth]{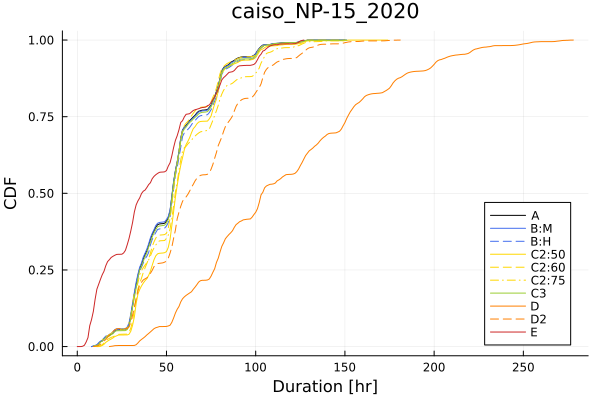}
\includegraphics[width=0.32\textwidth]{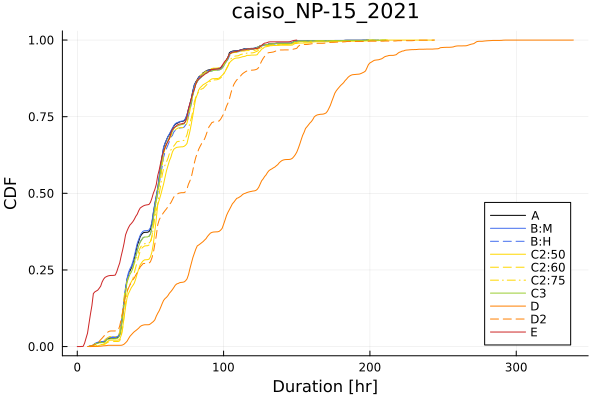}
\caption{ECDFs of energy storage duration for CAISO NP-15 zone.}
\end{figure}

\begin{figure}[h]
\centering
\includegraphics[width=0.32\textwidth]{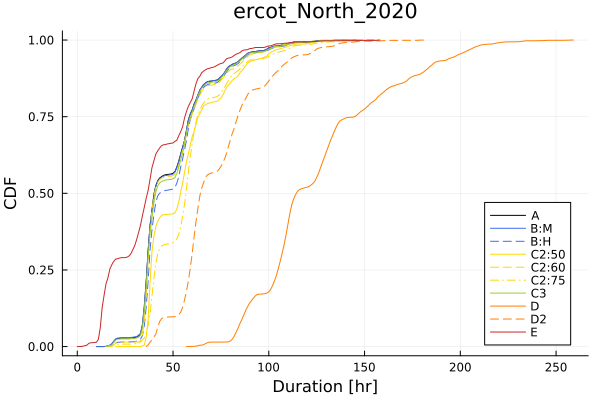}
\includegraphics[width=0.32\textwidth]{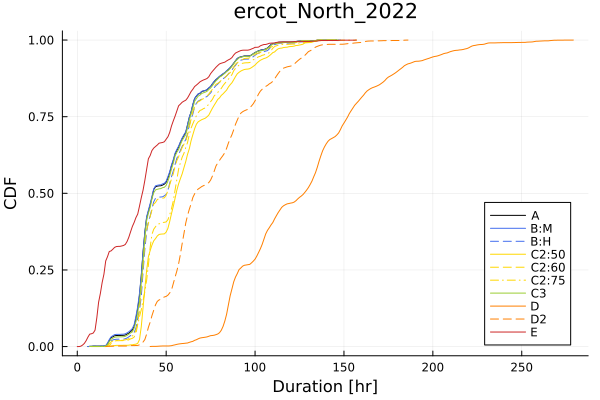}
\caption{ECDFs of energy storage duration for ERCOT North hub.}
\end{figure}

\begin{figure}[h]
\centering
\includegraphics[width=0.32\textwidth]{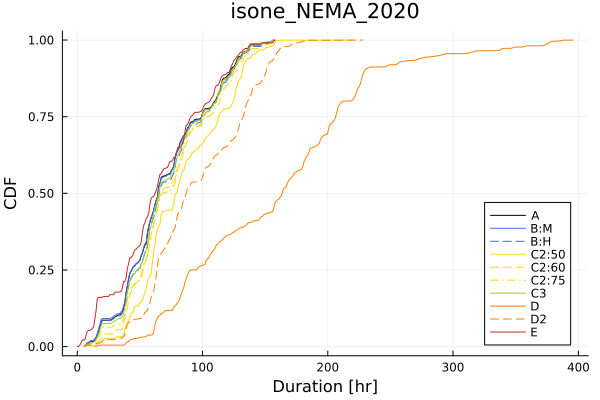}
\includegraphics[width=0.32\textwidth]{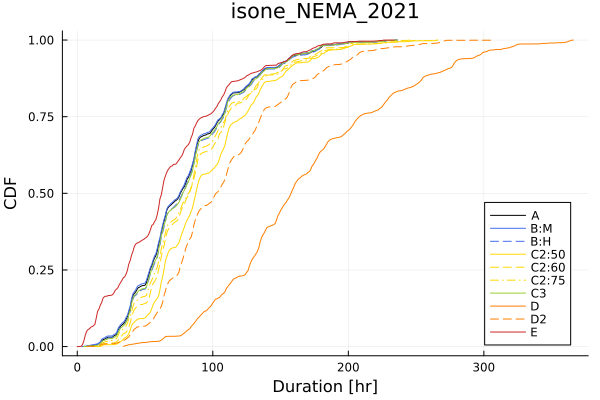}
\includegraphics[width=0.32\textwidth]{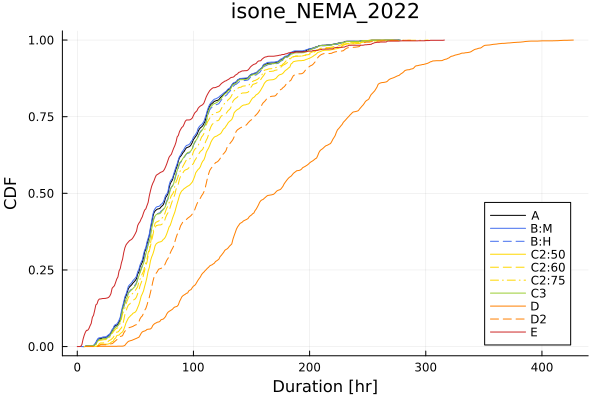}
\caption{ECDFs of energy storage duration for ISONE NEMA zone.}
\end{figure}

\begin{figure}[h]
\centering
\includegraphics[width=0.32\textwidth]{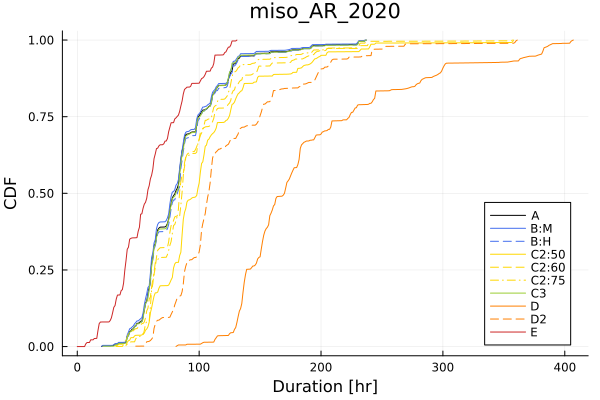}
\includegraphics[width=0.32\textwidth]{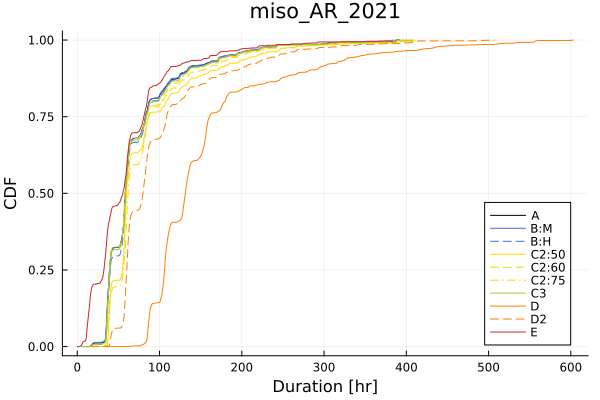}
\includegraphics[width=0.32\textwidth]{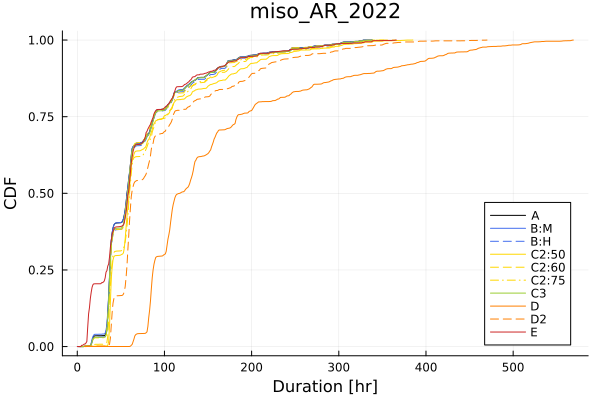}
\caption{ECDFs of energy storage duration for MISO AR hub.}
\end{figure}

\begin{figure}[h]
\centering
\includegraphics[width=0.32\textwidth]{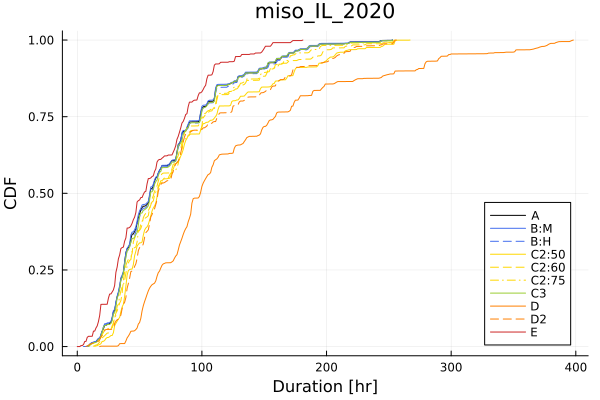}
\includegraphics[width=0.32\textwidth]{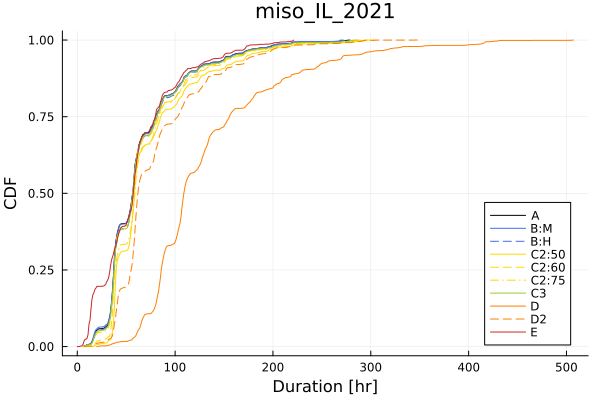}
\includegraphics[width=0.32\textwidth]{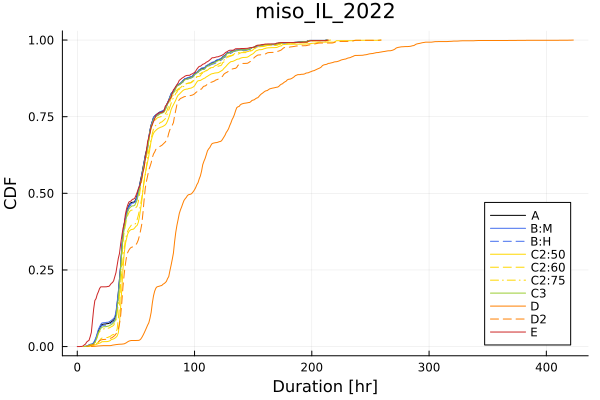}
\caption{ECDFs of energy storage duration for MISO IL hub.}
\end{figure}

\begin{figure}[h]
\centering
\includegraphics[width=0.32\textwidth]{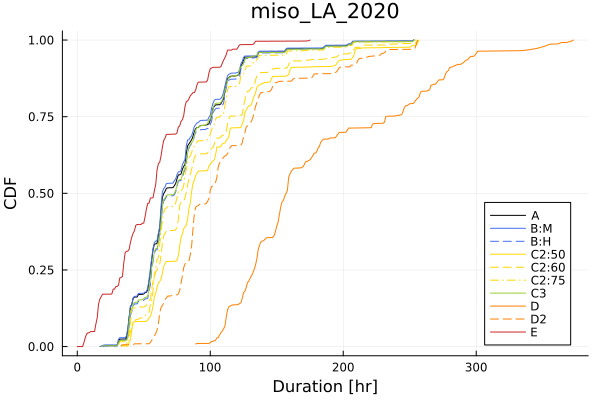}
\includegraphics[width=0.32\textwidth]{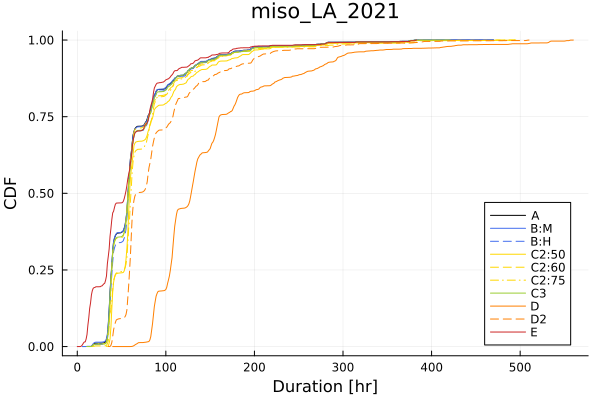}
\includegraphics[width=0.32\textwidth]{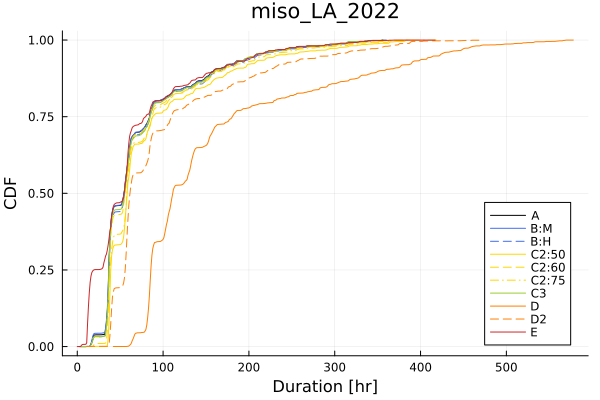}
\caption{ECDFs of energy storage duration for MISO LA hub.}
\end{figure}

\begin{figure}[h]
\centering
\includegraphics[width=0.32\textwidth]{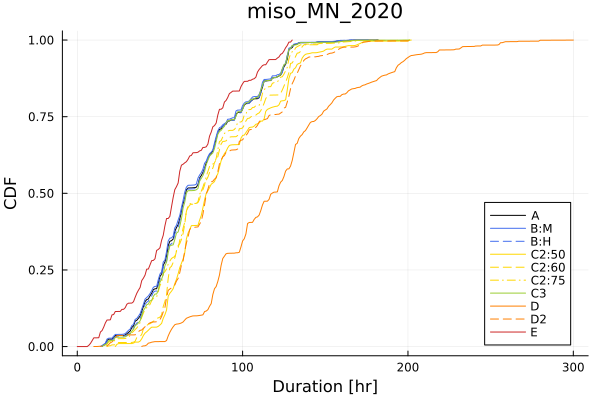}
\includegraphics[width=0.32\textwidth]{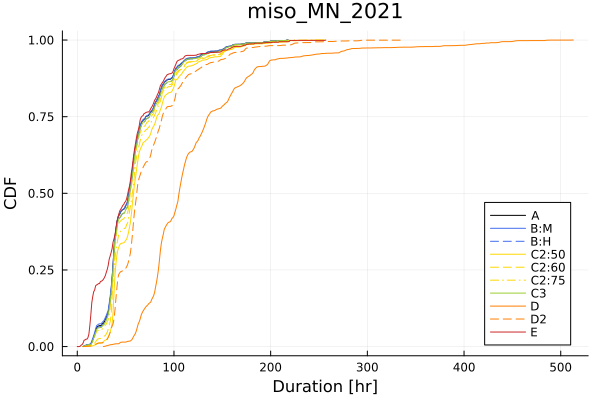}
\includegraphics[width=0.32\textwidth]{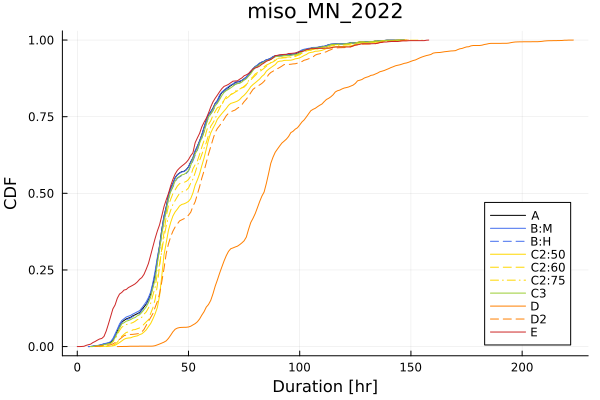}
\caption{ECDFs of energy storage duration for MISO MN hub.}
\end{figure}

\begin{figure}[h]
\centering
\includegraphics[width=0.32\textwidth]{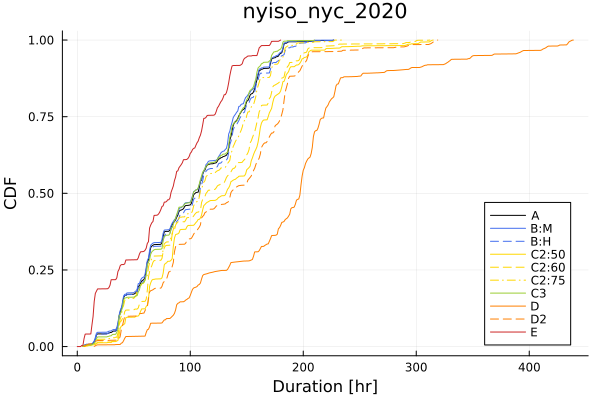}
\includegraphics[width=0.32\textwidth]{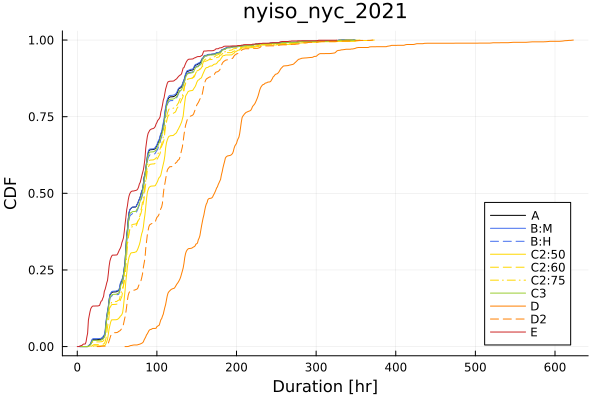}
\includegraphics[width=0.32\textwidth]{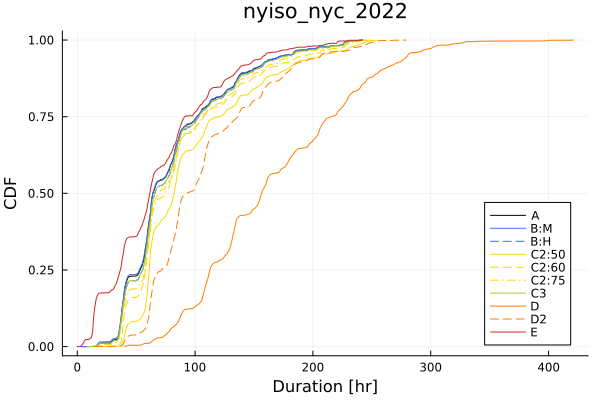}
\caption{ECDFs of energy storage duration for NYISO NYC zone.}
\end{figure}

\begin{figure}[h]
\centering
\includegraphics[width=0.32\textwidth]{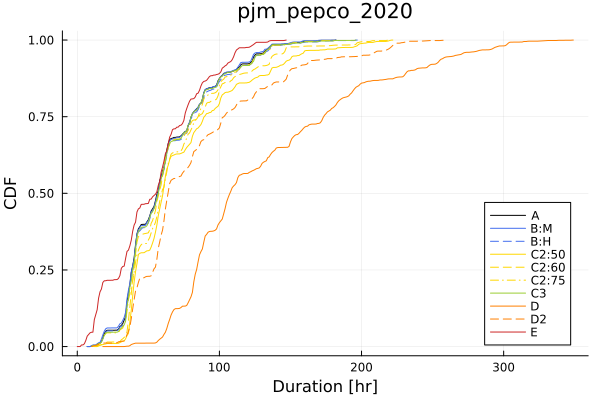}
\includegraphics[width=0.32\textwidth]{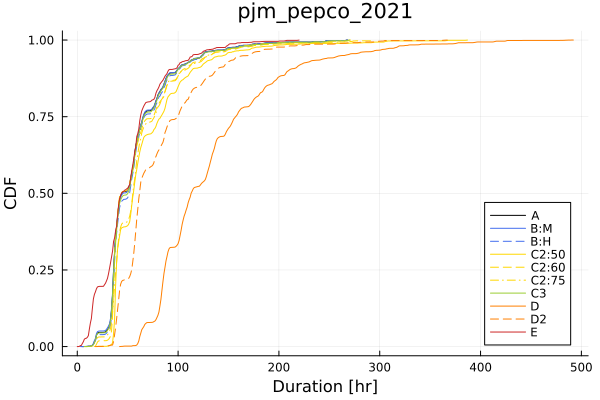}
\includegraphics[width=0.32\textwidth]{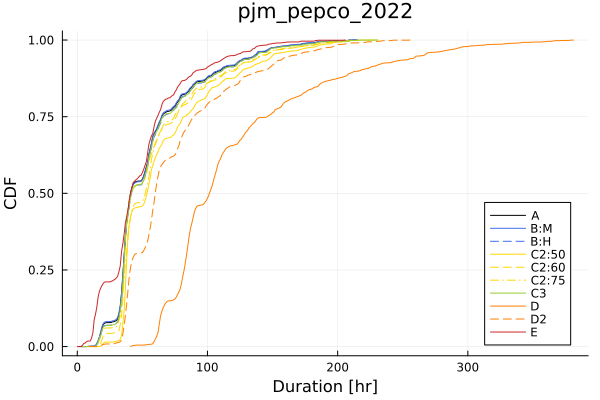}
\caption{ECDFs of energy storage duration for PJM Pepco zone.}
\end{figure}

\begin{figure}[h]
\centering
\includegraphics[width=0.32\textwidth]{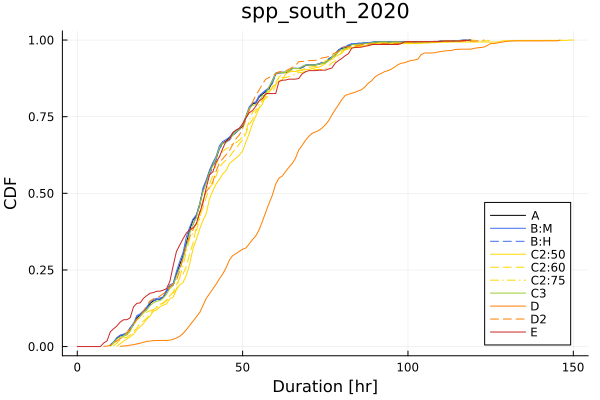}
\includegraphics[width=0.32\textwidth]{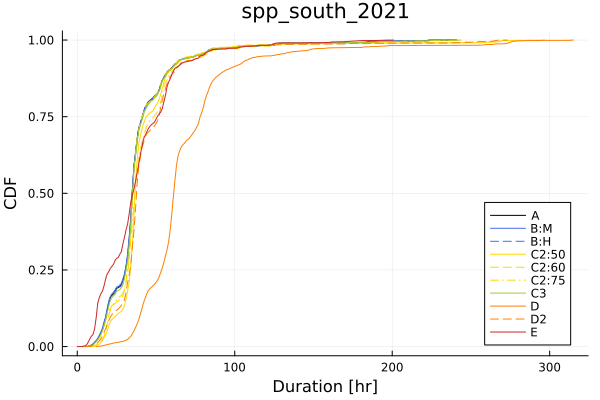}
\caption{ECDFs of energy storage duration for SPP South hub.}
\end{figure}

\clearpage

\section{Figures of merit for each price datase}\label{app6}
\begin{figure}[h]
\centering
\includegraphics[width=\textwidth]{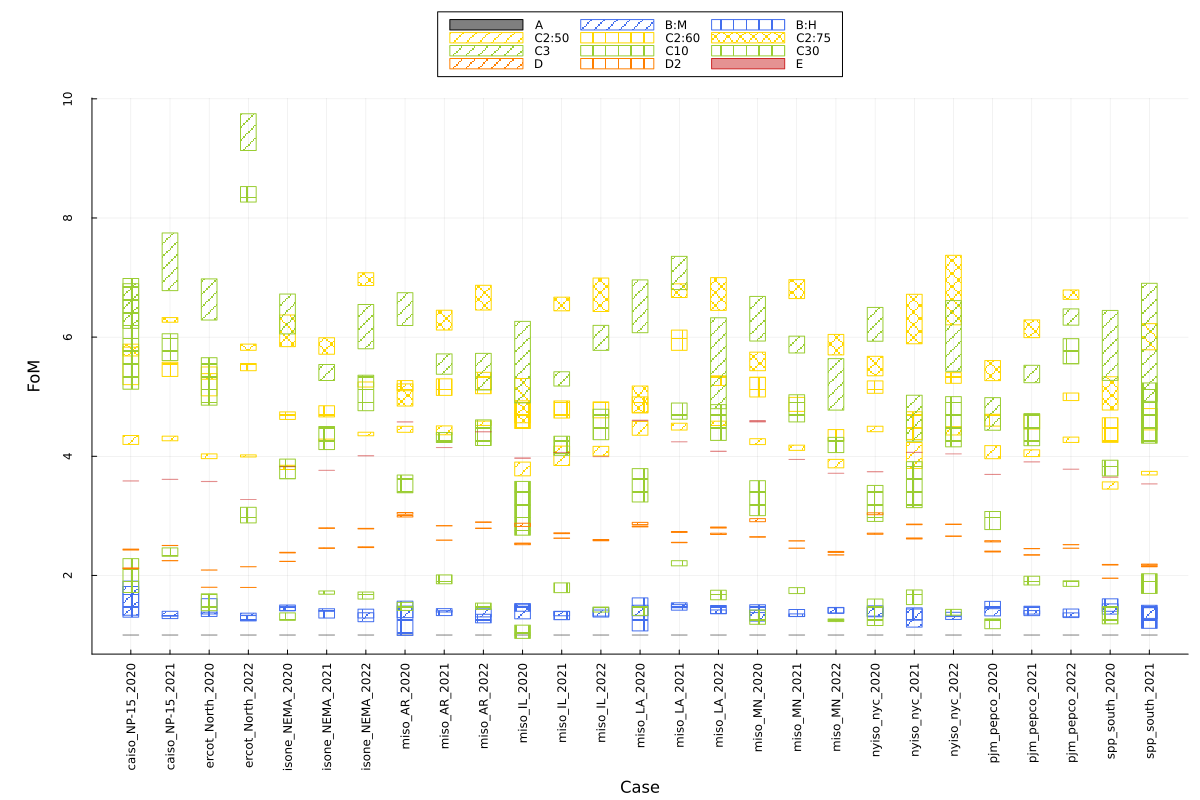}
\caption{FoMs for each price dataset with equal weighting on accuracy and computational efficiency.}\label{fig:fom1}
\end{figure}

\begin{figure}[h]
\centering
\includegraphics[width=\textwidth]{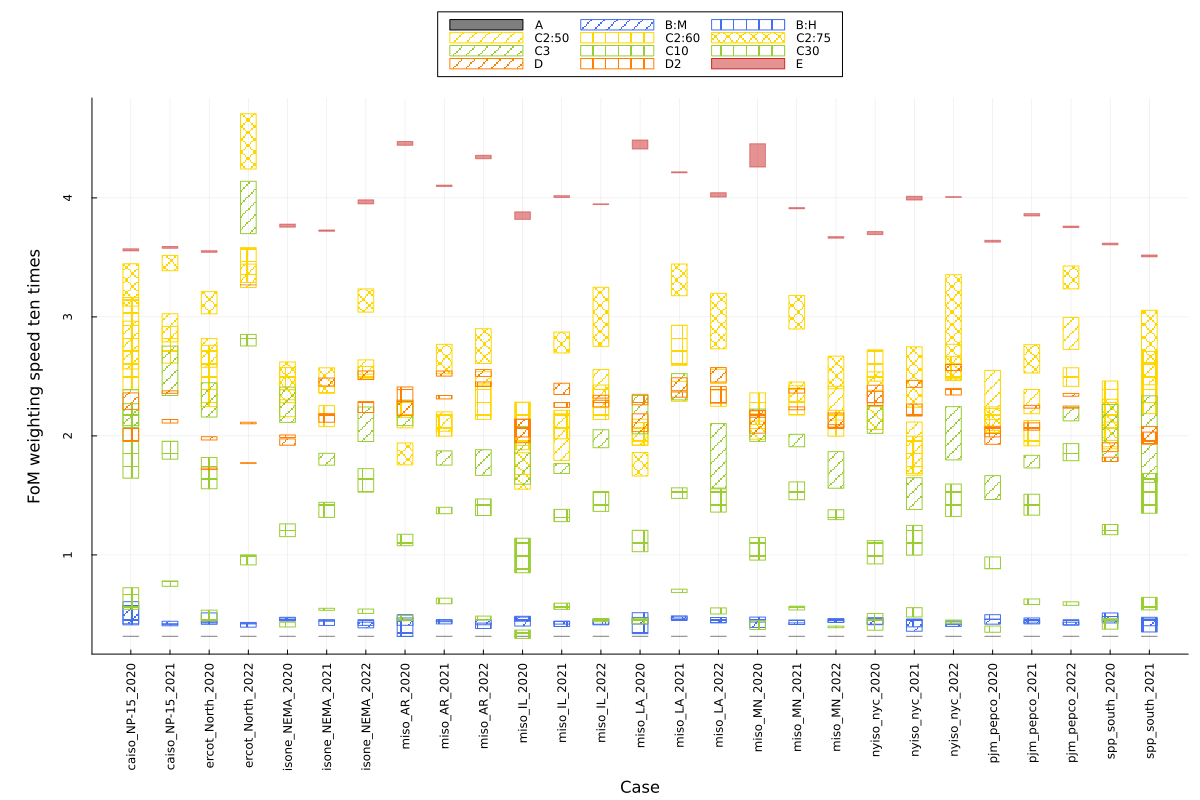}
\caption{FoMs for each price dataset with weighting computational efficiency ten times.}\label{fig:fom2}
\end{figure}

Figures of merit (FoM) for each price dataset are shown in this section, with min-max bars representing five runs. Fig.~\ref{fig:fom1} displays the FoM as described in Eq.~14, while Fig.~\ref{fig:fom2} presents the FoM with computational efficiency weighted ten times more.

\clearpage

\section{GenX charging and discharging operation comparison of Models A and E}\label{app7}
\begin{figure}[ht]
\centering
\includegraphics[width=\textwidth]{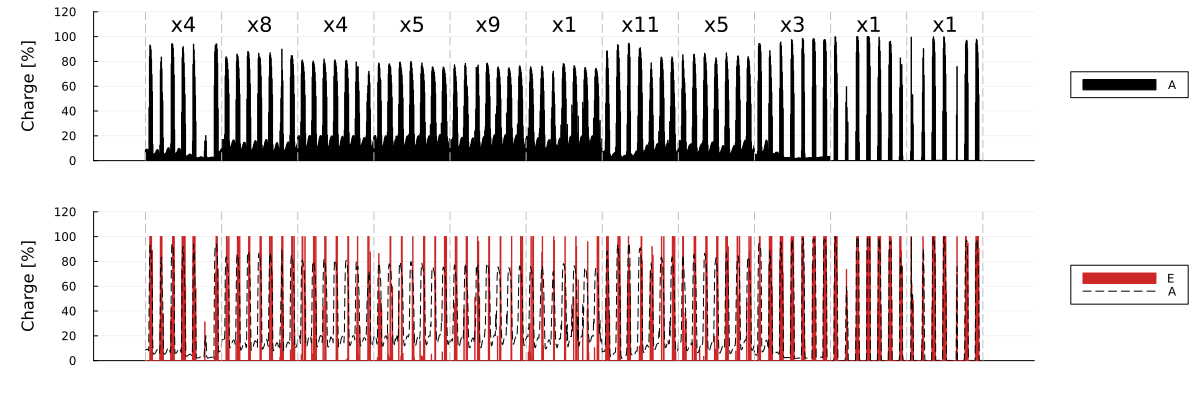}
\includegraphics[width=\textwidth]{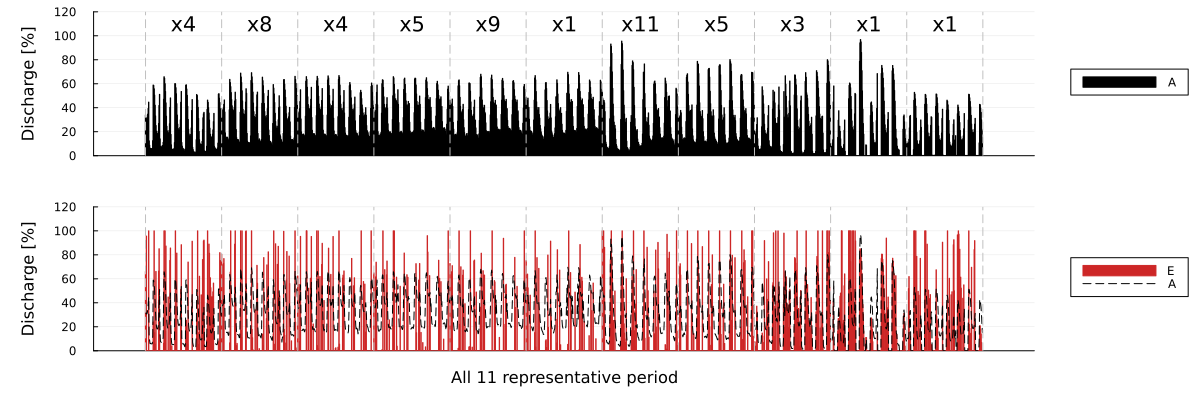}
\caption{Optimized charging and discharging patterns of Models A and E in GenX.}\label{fig:GenX_W}
\end{figure}

Figure~\ref{fig:GenX_W} highlights a distinct difference in the charging and discharging behaviors of Models A and E. Model A’s charging and discharging power is constrained by its capability function, while Model E operates at its maximum designed power for charging and discharging without regard to capability limitations.

\section{Alternative Model C variants formulation}\label{app8}
Model C variants can be alternatively formulated by using integer variables to choose one of the linear functions that approximate $\eta_{ch}^{B:M}(SoC)$ and $\eta_{dis}^{B:M}(SoC)$, as described in Eq.~\ref{eq:eta_C2_MILP}. This approach makes the variants mixed integer linear programs, and it can be implemented using logarithmic formulations, as suggested by \citet{huchette2023nonconvex}.
\begin{equation}
\begin{split}
\eta_{ch}^{C2:X}(SoC) &= \begin{cases} \frac{\eta_{ch}^{B:M}(X)-1}{X}SoC + 1, &SoC < X\%\\ \frac{\eta_{ch}^{B:M}(X)}{100-X}(100-SoC), &SoC \geq X\%\end{cases}\\
\eta_{dis}^{C2:X}(SoC) &= \begin{cases} \frac{\eta_{dis}^{B:M}(100-X)}{100-X}SoC, &SoC < 100-X\% \\ \frac{1 - \eta_{dis}^{B:M}(100-X)}{X}(SoC-100) + 1, &SoC \geq 100-X\%\end{cases}
\end{split}\label{eq:eta_C2_MILP}
\end{equation}

\begin{figure}[ht]
\centering
\includegraphics[width=0.7\textwidth]{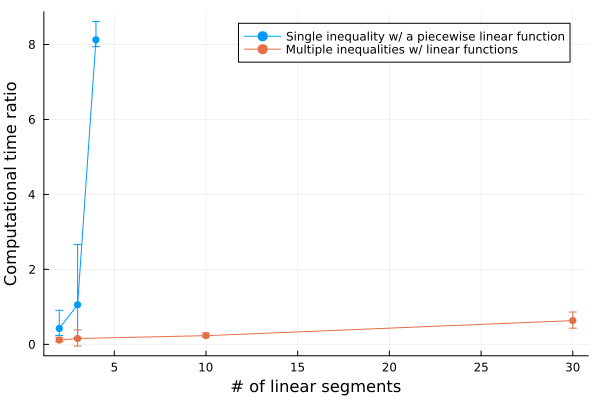}
\caption{Computational time ratio versus the number of linear segments for two approaches to formulate Model C variants: 1) single inequality with a piecewise linear function and 2) multiple inequalities with linear functions.}\label{fig:piecewise_multi}
\end{figure}

Figure~\ref{fig:piecewise_multi} shows how two Model C variant formulations scale with respect to the number of linear segments. All five runs of the Model C variants for 27 price data sets are used to evaluate the mean computational time ratio and the min-max range. Fig.~\ref{fig:piecewise_multi} clearly demonstrates that using multiple inequalities with linear functions scales better than using a single inequality with a piecewise linear function. When four linear segments are used with a piecewise linear function, the computational speed becomes even slower than Model A. Another challenge arises with the piecewise linear function formulation when coupled with GenX optimization. Due to its mixed-integer nature and the nonlinear programs required for capability functions, the optimization becomes a mixed-integer nonlinear program, which is significantly more complex compared to using multiple inequalities with linear functions. In these reasons, we used and recommend to use of the Model C variant formulation introduced in \ref{sec:2.3.3}.

\clearpage

\section{A computational performance observation with different Ipopt solvers}\label{app9}

\begin{figure}[ht]
\centering
\includegraphics[width=0.7\textwidth]{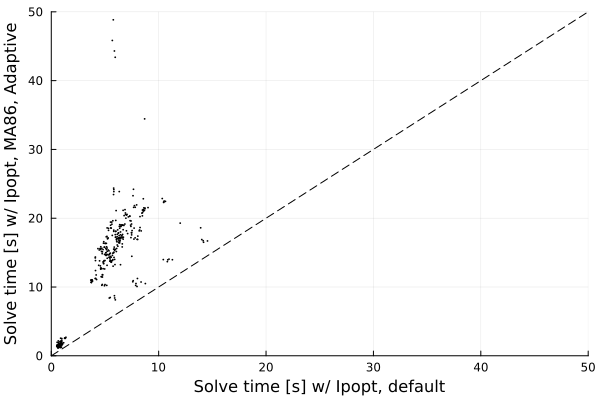}
\caption{A solve time comparison between Ipopt with the default solver and Ipopt with the MA86 linear solver and adaptive setting. All five runs of the 27 price data sets in the price-taker problem with Models A, B:M, and B:H are shown in the scatter plot.}\label{fig:mumps_ma86}
\end{figure}

Ipopt's default solver, MUMPS, performed faster in the first case study, while the MA86 linear solver, combined with setting the Ipopt parameter  ‘‘mu\_strategy’’ to ‘‘adaptive’’, was faster for GenX with more detailed models. Fig.~\ref{fig:mumps_ma86} shows the Ipopt solve times for all five runs of the 27 price data sets in the price-taker case study with Models A, B:M, and B:H. In most cases, the MA86 linear solver took longer than MUMPS to find the optimal solution. However, we observed that MUMPS slowed down exponentially when the optimization problem reached its memory limit, an issue the MA86 linear solver effectively addressed. We recommend using the default Ipopt setting for most users and switching to the MA86 linear solver with the adaptive parameter setting only when additional memory allocation is required for more complex investment modeling.

\clearpage

\section*{Data availability}
The GenX electricity system capacity expansion model is open-source and available on GitHub at \url{https://github.com/GenXProject/GenX}. All necessary inputs, result datasets, and source code for the PTES models, price-taker optimization, and the modified version of GenX used in this work are available on Zenodo at \url{https://doi.org/10.5281/zenodo.13900072}.

\section*{Acknowledgments}
This work was supported by the MIT Energy Initiative’s Future Energy Systems Center. The authors thank the Center for its funding and support.

\bibliographystyle{cas-model2-names}
\bibliography{references}

\end{document}